\documentclass[]{bmcart}

\usepackage{amsthm,amsmath}
\usepackage{bm}
\RequirePackage[authoryear]{natbib}
\RequirePackage{hyperref}
\hypersetup{
	colorlinks=true,
	linkcolor=black,
	filecolor=black,      
	urlcolor=black,
}
\usepackage[utf8]{inputenc} 
\usepackage{graphicx}
\graphicspath{{./Figs/}}
\usepackage{cancel}
\usepackage{booktabs}
\usepackage{url}

\startlocaldefs

\newcommand{\Bb}{{\boldsymbol{\mathnormal b}}}

\newcommand{\Be}{{\boldsymbol{\mathnormal e}}}
\newcommand{\Bf}{{\boldsymbol{\mathnormal f}}}

\newcommand{\Bl}{{\boldsymbol{\mathnormal l}}}

\newcommand{\Bn}{{\boldsymbol{\mathnormal n}}}

\newcommand{\Br}{{\boldsymbol{\mathnormal r}}}
\newcommand{\Bs}{{\pmb{\mathnormal s}}}
\newcommand{\Bt}{{\boldsymbol{\mathnormal t}}}
\newcommand{\Bu}{{\boldsymbol{\mathnormal u}}}
\newcommand{\Bv}{{\boldsymbol{\mathnormal v}}}

\newcommand{\BQ}{{\boldsymbol{\mathnormal Q}}}

\newcommand{\superscr}[1]{\ensuremath{{}^{\rm #1}}}

\newcommand{\Balpha }{\ensuremath{\boldsymbol\alpha}}

\newcommand{\Brho}{{\boldsymbol{\rho}}}

\newcommand{\Bve    }{\ensuremath{\boldsymbol\varepsilon}}

\newcommand{\vp     }{\ensuremath{\varphi}}
\newcommand{\Bbetapl}{\ensuremath{\boldsymbol\beta\superscr{pl}}}
\newcommand{\Bbetael}{\ensuremath{\boldsymbol\beta\superscr{el}}}

\newcommand{\rhot   }{\ensuremath{           \rho }}
\newcommand{\rhotot   }{\ensuremath{           \rho \superscr{tot}}}

\newcommand{\qt   }{\ensuremath{                q }}

\newcommand{\AO     }{{\ensuremath{   {\rho} }}}
\newcommand{\AI     }{{\ensuremath{   {\Brho} }}}

\newcommand{\whAIone}{\ensuremath{\widehat\rho_{1}}}  
\newcommand{\whAItwo}{\ensuremath{\widehat\rho_{2}} } 
\newcommand{\AIone}{\ensuremath{\rho_{1}^{(1)}} } 
\newcommand{\AII     }{{\ensuremath{ {\Brho}^{(2)}}}}
\newcommand{\AIIoneone}{\ensuremath{\rho_{11}^{(2)}} } 
\newcommand{\AIItwotwo}{\ensuremath{\rho_{22}^{(2)}} } 
\newcommand{\AIIonetwo}{\ensuremath{\rho_{12}^{(2)}} } 
\newcommand{\AIII     }{{\ensuremath{   {\Brho} \superscr{(3)}}}}
\newcommand{\QI     }{{\ensuremath{   {\BQ}}}}
\newcommand{\QII     }{{\ensuremath{   {\BQ} \superscr{(2)}}}}

\newcommand{\lk}{\ensuremath{{\Bl}^{\Brho}}}
\newcommand{\lkone}{\ensuremath{{l}_1}}
\newcommand{\lktwo}{\ensuremath{{l}_2}}
\newcommand{\lkp}{\ensuremath{{\Bl}^{\Brho\perp}}}
\newcommand{\CDDO}{{\ensuremath{{\rm CDD}^{(0)}}}}
\newcommand{\CDDI}{{\ensuremath{{\rm CDD}^{(1)}}}}
\newcommand{\CDDII}{{\ensuremath{{\rm CDD}^{(2)}}}}

\newcommand{\MI}{{\ensuremath{{M} \superscr{(1)}}}}
\newcommand{\MII}{{\ensuremath{{M} \superscr{(2)}}}}
\newcommand{\MIII}{{\ensuremath{{M} \superscr{(3)}}}}

\newcommand{\td}{{\ensuremath{\textrm{d}}}}

\newcommand{\rphi}{{(\Br,\vp)}}        
        
\newcommand{\wh}[1]{{\widehat{#1}}}

\newcommand{\figref}[1]{Fig.~\ref{#1}}
\newcommand{\Figref}[1]{Fig.~\ref{#1}}
\newcommand{\secref}[1]{Section~\ref{#1}}

\newcommand{\appref}[1]{Appendix~\ref{#1}}

\newcommand \MZ [1] {\bgroup\noindent[\textcolor{blue}{\textbf{MZ}: #1}]\egroup\ignorespacesafterend}

\newcommand \Mdel [1] {\bgroup\noindent[\textcolor{red}{\textbf{Mdel}: #1}]\egroup\ignorespacesafterend}
\endlocaldefs

\newcommand \Madd [1] {\bgroup\noindent[\textcolor{blue}{\textbf{Madd}: #1}]\egroup\ignorespacesafterend}
\endlocaldefs

\begin{document}

\begin{frontmatter}

\begin{fmbox}
\dochead{Research}


\title{Annihilation and sources in continuum dislocation dynamics}


\author[
   addressref={aff1},                   
   email={mehran.monavari@fau.de}   
   corref={aff1},                       
]{\inits{MM}\fnm{Mehran} \snm{Monavari}}
\author[
   addressref={aff1,aff2},
   email={michael.Zaiser@fau.de}
]{\inits{MZ}\fnm{Michael} \snm{Zaiser}}


\address[id=aff1]{
  \orgname{Institute of Materials Simulation (WW8), Friedrich-Alexander University Erlangen-N\"urnberg (FAU)}, 
  \street{Dr.-Mack-Str. 77},                     %
  \postcode{90762}                                
  \city{F\"urth},                              
  \cny{Germany}                                    
}
\address[id=aff2]{
  \orgname{Department of Engineering and Mechanics, Southwest Jiaotong University}, 
  \city{Chengdu}
  \cny{P.R. China}                                    
}


\begin{artnotes}
\end{artnotes}

\end{fmbox}


\begin{abstractbox}

\begin{abstract} 
Continuum dislocation dynamics (CDD) aims at representing the evolution of systems of curved and connected dislocation lines in terms of density-like field variables. Here we discuss how the processes of dislocation multiplication and annihilation can be described within such a framework. We show that both processes are associated with changes in the volume density of dislocation loops: dislocation annihilation needs to be envisaged in terms of the merging of dislocation loops, while conversely dislocation multiplication is associated with the generation of new loops. Both findings point towards the importance of including the volume density of loops (or 'curvature density') as an additional field variable into continuum models of dislocation density evolution. We explicitly show how this density is affected by loop mergers and loop generation. The equations which result for the lowest order CDD theory allow us, after spatial averaging and under the assumption of unidirectional deformation, to recover the classical theory of Kocks and Mecking for the early stages of work hardening.
\end{abstract}




\end{abstractbox}
%

\end{frontmatter}



\section[Introduction]{Introduction}
Since the discovery of dislocations as carriers of plastic deformation, developing a continuum theory for motion and interaction of dislocations has been  a challenging task. Such a theory should address two interrelated problems: how to represent in a continuum setting the motion of dislocations, hence the kinematics of curved and connected lines, and how to capture dislocation interactions.

The classical continuum theory of dislocation (CCT) systems dates back to \citet{Kroener58_Book} and \citet{Nye53_AM}. This theory describes the dislocation system in terms of a rank-2 tensor field $\Balpha$ defined as the curl of the plastic distortion, $\Balpha =  \nabla \times \Bbetapl$. The rate of the plastic distortion due to the evolution of the dislocation density tensor reads $\partial_t \Bbetapl=\Bv \times \Balpha$ where the dislocation velocity vector $\Bv$ is defined on the dislocation lines. The time evolution of $\Balpha$  becomes \citep{Mura63_PM}
\begin{align}
\partial_t \Balpha =  \nabla \times [\Bv \times \Balpha].
\end{align}
This fundamental setting provided by the classical continuum theory of dislocation systems has, over the past two decades, inspired many models \citep[e.g.][]{Sedlacek03_PM, Xiang09_JMPS, Zhu15_JMPS}. Irrespective of the specific formulation, a main characteristic of the CCT is that, in each elementary volume, the dislocation tensor can measure only the minimum amount of dislocations which are necessary for geometrical compatibility of plastic distortion (`geometrically necessary' dislocations (GND)). If additional dislocations of zero net Burgers vector are present, CCT is bound to be incomplete as a plasticity theory because the `redundant' dislocations contribute to the averaged plastic strain rate and this contribution must be accounted for. Conversely we can state that CCT is working perfectly whenever `redundant' dislocations are physically absent. This condition is of course fulfilled for particular geometrical configurations, but in general cases it can be only met if the linear dimension of the elementary volume of a simulation falls below the distance over which dislocations spontaneously react and annihilate, such that 'redundant' dislocations cannot physically exist on the scale of the simulation. This simple observation demonstrates the close connection between the problem of averaging and the problem of annihilation - a connection which we will further investigate in detail in Section 3 of the present paper. 

From the above argument we see that one method to deal with the averaging problem is to remain faithful to the CCT framework and simply use a very high spatial resolution.  We mention, in particular, the recent formulation by El-Azab which incorporates statistical phenomena such as cross-slip \citep{Xia15_MSMSE} and time averaging \citep{Xia16_MSMSE} and has shown promising results in modelling dislocation pattern formation. This formulation is based upon a decomposition of the tensor $\Balpha$ into contributions of dislocations from the different slip systems $\varsigma$ in the form $\Balpha = \sum_{\varsigma}\Brho^{\varsigma}\otimes\Bb^{\varsigma}$ where $\Bb^{\varsigma}$ is the Burgers vector of dislocations on slip system $\varsigma$ and the dislocation density vector $\Brho^{\varsigma}$ of these dislocations points in their local line direction. Accordingly, the evolution of the dislocation density tensor is written as $\partial_t \Balpha = \sum_{\varsigma}\partial_t \Brho^{\varsigma}\otimes\Bb^{\varsigma}$ with $\partial_t \Brho^{\varsigma} = \nabla \times [\Bv^{\varsigma} \times \Brho^{\varsigma}]$ where the dislocation velocities $\Bv^{\varsigma}$ are again slip system specific.  We consider a decomposition of the dislocation density tensor into slip system specific tensors as indispensable for connecting continuum crystal plasticity to dislocation physics: it is otherwise impossible to relate the dislocation velocity $\Bv$ in a meaningful manner to the physical processes controlling dislocation glide and climb, as the glide and climb directions evidently depend on the respective slip system. We therefore use a description of the dislocation system by slip system specific dislocation density vectors as the starting point of our subsequent discussion. 

CCT formulated in terms of slip system specific dislocation density vectors with single-dislocation resolution is a complete and kinematically exact plasticity theory but, as the physical annihilation distance of dislocations is of the order of a few nanometers, its numerical implementation may need more rather than less degrees of freedom compared to a discrete dislocation dynamics model. There are nevertheless good reasons to adopt such a formulation: Density based formulations allow us to use spatio-temporally smoothed velocity fields which reduce the intermittency of dislocation motion in discrete simulations. Even more than long-range interactions and complex kinematics, the extreme intermittency of dislocation motion and the resulting numerical stiffness of the simulations is a main factor that makes discrete dislocation dynamics simulations computationally very expensive. Furthermore, in CCT, the elementary volume of the simulation acts as a reaction volume and thus annihilation does not need any special treatment.  

Moving from the micro- to the macroscale requires the use of elementary volumes that significantly exceed the annihilation distance of dislocations. Averaging operations are then needed which can account for the presence of 'redundant' dislocations. Some continuum theories try to resolve the averaging problem by describing the microstructure by multiple dislocation density fields which each represent a specific dislocation orientation $\vp$ on a slip system $\varsigma$. Accordingly, all dislocations of such a partial population move in the same direction with the same local velocity $\Bv_{\vp}^{\varsigma}$ such that $\langle\partial_t \Brho_{\vp}^{\varsigma}\rangle \approx \nabla \times [\Bv_{\vp}^{\varsigma} \times \langle\rho_{\vp}^{\varsigma}\rangle]$. Along this line Groma, Zaiser and co-workers \citep{Groma97_PRB, Zaiser01_PRB, Groma03_AM} developed  statistical approaches for evolution of 2D systems of straight, positive and negative edge dislocations. Inspired by such 2D models, \citet{Arsenlis04_JMPS,Reuber14_AM,Leung15_MSMSE} developed 3D models by considering additional orientations. However, extending the 2D approach to 3D systems where connected and curved dislocation lines can move perpendicular to their line direction while remaining topologically connected is not straightforward, and most models use, for coupling the motion of dislocations of different orientations, simplified kinematic rules that cannot in general guarantee  dislocation connectivity (see \citet{Monavari16_JMPS} for a detailed discussion).  

The third line takes a mathematically rigorous approach towards averaged, density-based  representation of generic 3D systems of curved dislocation lines based on the idea of  envisaging dislocations in a higher dimensional phase space where densities carry additional information about their line orientation and curvature in terms of continuous orientation variables $\vp$ \citep{Hochrainer06_Phd,Hochrainer07_PM}. In this phase space, the microstructure is described by dislocation orientation distribution functions (DODF) $\rho\rphi$. Tracking the evolution of a higher dimensional $\rho\rphi$ can be a numerically challenging task. Continuum dislocation dynamics (CDD) estimates the evolution of the DODF in terms of its alignment tensor expansion series \citep{Hochrainer15_PM}. The components of the dislocation density alignment tensors can be envisaged as density-like fields which contain more and more detailed information about the orientation distribution of dislocations. CDD has been used to simulate various phenomena including dislocation patterning \citep{Sandfeld15_MSMSEa,Wu17_arxiv} and co-evolution of phase and dislocation microstructure \citep{Wu17_IJP}. The formulation in terms of alignment tensors has proven particularly versatile since one can formulate the elastic energy functional of the dislocation system in terms of dislocation density alignment tensors \citep{Zaiser15_PRB} and then use this functional to derive the dislocation velocity in a thermodynamically consistent manner \citep{Hochrainer16_JMPS}. 

Alignment tensor based CDD at present suffers from an important limitation: While the total dislocation density changes due to elongation or shrinkage of dislocation loops, the number of loops is a conserved quantity. This leads to unrealistic dislocation starvation and hardening behaviour for bulk crystals \citep{Monavari14_MRSSP}. The goal of the present paper is to incorporate into the CDD theory mechanisms which change the number of dislocation loops by accounting for the merger of loops consequent to local annihilation of dislocation segments from different loops and for the formation of loops by operation of sources. First we revisit the hierarchical evolution equations of CDD. Then we introduce a kinematic model to describe the annihilation of dislocations in higher dimensional phase space. We calculate the annihilation rate for the variables of the lowest-order CDD theories. Then we introduce models for incorporating activation of Frank-Read sources, cross slip sources and glissile junctions into CDD. We demonstrate that by incorporating of annihilation (loop merger) and sources (loop generation) into CDD, even a lowest-order CDD formulation can predict the first 3 stages of work hardening. 

\section[CDD]{Continuum Dislocation Dynamics}

\subsection{Conventions and  notations}
We describe the kinematics of the deforming body by a displacement vector field $\Bu$. Considering linearised kinematics of small deformations we use an additive decomposition of the corresponding deformation gradient into elastic and plastic parts: $\nabla \Bu= \Bbetael + \Bbetapl$. 
Dislocations of Burgers vectors $\Bb^{\varsigma}$are assumed to move only by glide (unless stated otherwise) and are therefore confined to their slip planes with slip plane normal vectors $\Bn^{\varsigma}$. This motion generates a plastic shear $\gamma^{\varsigma}$ in the direction of the unit slip vector $\Bb^{\varsigma}/b$ where $b$ is the modulus of $\Bb^{\varsigma}$. We use the following sign convention: A dislocation loop which expands under positive resolved shear stress  is called a positive loop, the corresponding dislocation density vector $\Brho^{\varsigma}$ points in counter-clockwise direction with respect to the slip plane normal $\Bn^{\varsigma}$. Summing the plastic shear tensors of all slip systems gives the plastic distortion: $\Bbetapl=\sum_{\varsigma}\gamma^{\varsigma} \Bn^{\varsigma} \otimes \Bb^{\varsigma}/b$. 

On the slip system level, without loss of generality, we use a Cartesian coordinate system with unit vectors $\Be_1^{\varsigma}= \Bb^{\varsigma}/b$, $\Be_3= \Bn^{\varsigma}$ and $\Be_2 = \Bn^{\varsigma}\times\Bs^{\varsigma}$. A slip system specific Levi-Civita tensor $\Bve^{\varsigma}$ with coordinates $\varepsilon_{ij}^{\varsigma}$ is constructed by contracting the fully antisymmetric Levi-Civita operator with the slip plane normal, $\varepsilon_{ij}^{\varsigma}=\varepsilon_{ikj}n_k^{\varsigma}$. The operation $\Bt.\Bve^{\varsigma} =: \Bt_{\perp}$ then rotates a vector $\Bt$ on the slip plane clockwise by $90^\circ$ around $\Bn^{\varsigma}$. In the following we drop, for brevity, the superscript $\varsigma$ as long as definitions and calculations pertaining to a single slip system are concerned. 

The quantity which is fundamental to density based crystal plasticity models is the slip system specific dislocation density vector $\Brho$. The modulus of this vector defines a scalar density $\rho = |\Brho|$ and the unit vector $\Bl = \Brho/\rho$ gives the local dislocation direction. The $m^{{\rm th}}$ order power tensor of $\Bl$ is defined by the recursion relation $\Bl^{\otimes 1} =\Bl$, $\Bl^{\otimes m+1} =\Bl^{\otimes m}\otimes\Bl$. In the slip system coordinate system, $\Bl$ can be expressed in terms of the orientation angle $\vp$ between the line tangent and the slip direction as $\Bl(\varphi)=\cos(\varphi)\Be_1 + \sin(\varphi)\Be_2$. When considering volume elements containing dislocations of many orientations, or ensembles of dislocation systems where the same material point may in different realizations be 
occupied by dislocations of different orientations, we express the local statistics of dislocation orientations in terms of the probability density function $p_{\Br}(\vp)$ of the orientation angle $\vp$ within a volume element located at $\Br$. We denote $p_{\Br}(\vp)$ as the local dislocation orientation distribution function (DODF). The DODF is completely determined by the set of moments $\langle \vp^n \rangle_{\Br}$ but also by the expectation values of the  power tensor series $\langle \Bl^n \rangle_{\Br}$. The latter quantities turn out to be particularly useful for setting up a kinematic theory. Specifically, the so-called dislocation density alignment tensors
\begin{align}
\label{eq:An}
\Brho^{(n)}(\Br) :=  \rho \langle \Bl^{\otimes n} \rangle_{\Br} = \rho \oint \,\!\!p_{\Br}(\vp) 
{\Bl(\vp)}^{ \otimes n}
\,{\rm d}\vp.
\end{align}
turn out to be suitable field variables for constructing a statistically averaged theory of dislocation kinematics. Components of the k-th order alignment tensor $\Brho^{(k)}(\Br)$ are denoted $\rho_{{a_1}\dots{a_k}}$. $\wh \Brho^{(n)}(\Br)=\Brho^{(n)}(\Br)/\rhot$ denotes normalization of an alignment tensor by dividing it by the total dislocation density; this quantity equals the DODF-average of the $n$th order power tensor of $\Bl$. ${\rm Tr}(\bullet)$ gives the trace of a symmetric alignment tensor by summation over any two indices. The symmetric part of a tensor is denoted by $[\bullet]_{\rm sym}$. The time derivative of the quantity $x$ is denoted by $\partial_t(x)$ or by $\dot x$.

\subsection{Kinematic equations of Continuum Dislocation Dynamics (CDD) theory} \label{subsec:CDD}

\citet{Hochrainer15_PM} derives the hierarchy of evolution equations for dislocation density alignment tensors by first generalizing the CCT dislocation density tensor to a higher dimensional space which is the direct product of the 3D Euclidean space and the space of line directions (second-order dislocation density tensor, SODT). Kinematic evolution equations for the SODT are obtained in the framework of the calculus of differential forms and then used to derive equations for alignment tensors by spatial projection. For a general and comprehensive treatment we refer the reader to \citet{Hochrainer15_PM}. Here we motivate the same equations in terms of probabilistic averaging over single-valued dislocation density fields, considering the case of deformation by dislocation glide. 

We start from the slip system specific Mura equation in the form 
\begin{equation}
\partial_t \Brho = \nabla \times [\Bv \times \Brho].
\label{eq:Mura1}
\end{equation}
where for simplicity of notation we drop the slip system specific superscript $\varsigma$ and we assume that the spatial resolution is sufficiently high such that the dislocation line orientation $\Bl$ is uniquely defined in each spatial point. If deformation occurs by crystallographic slip, then the dislocation velocity vector must in this case have the local direction $\Be_v = \Bl \times \Bn = \Brho\times \Bn/\rho$. This implies that the Mura equation is {\em kinematically non-linear}: writing the right-hand side out we get
\begin{equation}
\partial_t \Brho = \nabla \times [\Bl \times \Bn \times \Brho v] = \nabla \times [\frac{\Brho \times \Bn \times \Brho}{|\Brho|} v].
\label{eq:Mura2}
\end{equation}
where the velocity magnitude $v$ depends on the local stress state and possibly on dislocation inertia. This equation is non-linear
even if the dislocation velocity $v$ does not depend on $\Brho$, and this inherent kinematic non-linearity 
makes the equation difficult to average. To obtain an equation which is linear in a dislocation density variable and therefore can
be averaged in a straightforward manner (i.e., by simply replacing the dislocation density variable by its average) is, however, possible: 
We note that $\Brho = \Bl \rho$ and $\nabla \times \Bl \times \Bn 
\times \Bl = -\Bve.\nabla$, hence
\begin{eqnarray}
\partial_t \Brho = -\Bve\cdot\nabla(\rho v).
\label{eq:Murarewrite}
\end{eqnarray}
In addition we find because of $\Brho \otimes \Bb = \nabla\times \Bbetapl$ that the plastic strain rate and the shear strain rate on the considered slip system fulfil the Orowan equation 
\begin{equation}
\partial_t \Bbetapl = [\Bn \otimes \Bb] \rho v = [\Bn \otimes \Bs] \partial_t \gamma \quad,\quad
\partial_t \gamma = \rho b v.
\label{eq:Strainrate}
\end{equation}
We now need to derive an equation for the scalar density $\rho$. This is straightforward: we use that $\rho^2 = \Brho.\Brho$, hence $\partial_t \rho = (\Brho/\rho) \cdot \partial_t \Brho$. After a few algebraic manipulations we obtain
\begin{equation}
\partial_t \rho = \nabla \cdot (\Bve \cdot \Brho v) + q v
\label{eq:rho}
\end{equation}
where we introduced the notation
\begin{equation}
q := - \rho (\nabla\cdot\Bve\cdot\Bl) .
\end{equation}
To interpret this new variable we observe that $k= - \nabla\cdot\Bve\cdot \Bl = \partial_1 l_2 - \partial_2 l_1$ is the curvature of the unit vector field $\Bl$, i.e. the reciprocal radius of curvature of the dislocation line \citep{Theisel95_Rostock}. Hence the product $q = \rho k$ can be called a curvature density. Integration of $q$ over a large volume $V$ yields the number of loops contained in $V$, hence, $q$ may also be envisaged as a loop density. 

The quantity $q$ defines a new independent variable. Its evolution equation is obtained from those of $\rho$ and $\Bl = \Brho/\rho$. After some algebra we get
\begin{equation}
\partial_t q = \nabla \cdot (v\BQ -  \Brho^{(2)}\cdot\nabla v ) .
\label{eq:q}
\end{equation}
where we have taken care to write the right-hand side in a form that contains density- and curvature-density like variables in a linear manner.
As a consequence, on the right hand side appears a second order tensor $\Brho^{(2)} = \rho \Bl \otimes \Bl = \Brho \otimes \Brho/\rho$. By using the fact that $\Brho$ is divergence-free, $\nabla.\Brho = \nabla.(\rho \Bl) = 0$, we can show that the vector $\BQ = q \Bve\cdot\Bl$ derives from this tensor according to $\BQ = \nabla.\Brho^{(2)}$. 

We thus find that the equation for the curvature density $q$ contains a rank-2 tensor which can be envisaged as the normalized power tensor of the dislocation density vector. On the next higher level, we realize that the equation for $\Brho^{(2)}$ contains higher-order curvature tensors, leading to an infinite hierarchy of equations given in full by 
\begin{align}
\label{eq:dA0dt}
\partial_t\rho &=\nabla \cdot(v \Bve\cdot\AI )+v, \\
\label{eq:dAkdt}
\partial_t\Brho^{(n)} &=\left[-\Bve\cdot\nabla (v \Brho^{(n-1)} )+(n-1)v\BQ^{(n)}-(n-1) \Bve\cdot\Brho^{(n+1)}\cdot\nabla v \right]_{\text{sym}},\\
\label{eq:dqtdt}
\partial_t\qt &=\nabla \cdot(v\BQ^{(1)} -  \Brho^{(2)}\cdot\nabla v ) ,
\end{align}
 where $\BQ^{(n)}$ are auxiliary symmetric curvature tensors defined as 
 \begin{align}
 \label{eq:Qn}
 \BQ^{(n)} &= \qt\Bve \cdot \Bl \otimes\Bve \cdot \Bl \otimes
 \Bl^{ \otimes n-2}. 
 \end{align}
So far, we have simply re-written the single, kinematically non-linear Mura equation in terms of an equivalent infinite hierarchy of kinematically linear equations for an infinite set of dislocation density-like and curvature-density like variables. The idea behind
this approach becomes evident as soon as we proceed to perform averages over volumes containing dislocations of many orientations, or
over ensembles where in different realizations the same spatial point may be occupied by dislocations of different orientations. The fact
that our equations are linear in the density-like variables allows us to average them over the DODF $p(\vp)$ while {\em retaining the 
functional form} of the equations. The averaging simply replaces the normalized power tensors of the dislocation density vector by their DODF-weighted averages, i.e., by the respective dislocation density alignment tensors: 
\begin{align}
\Brho^{(n)}(\Br) \to \oint p_{\Br}(\vp) \Brho^{(n)}(\Br) {\rm d} \vp
\end{align}
and similarly
\begin{align}
\BQ^{(n)}(\Br) \to \oint p_{\Br}(\vp) \BQ^{(n)}(\Br) {\rm d} \vp.
\end{align}

The problem remains that we now need to close the infinite hierarchy of evolution equations of the alignment tensors. A theory that uses alignment tensors of order $k$ can be completely specified by the evolution equation of $q$ together with the equations for the $\Brho^{(k-1)}$ and $\Brho^{(k)}$ tensors (lower order tensors can be obtained from these by contraction). To close the theory, the tensor $\Brho^{(k+1)}$ needs to be approximated in terms of lower order tensors.  A systematic approach for deriving closure approximations was proposed by Monavari \citep{Monavari16_JMPS}. The fundamental idea is to use the Maximum Information Entropy Principle (MIEP) in order to estimate the DODF based upon the information contained in alignment tensors up to order $k$, and then use the estimated DODF to evaluate, from Eq. (\ref{eq:An}), the missing alignment tensor $\Brho^{(k+1)}$. This allows to close the evolution equations at any desired level. 

For example, closing the theory at zeroth order is tantamount to assuming a uniform DODF for which the corresponding closure relation reads $\Brho^{(1)} \approx 0$. The evolution equations of \CDDO\ then are simply  
\begin{align}
\label{eq:dA0dtO}
\partial_t\rhot &= v q \\
\label{eq:dqtdtO}
\partial_t q &= 0
\end{align}
These equations represent the expansion of a system consisting of a constant number of loops. In \secref{sec:CDDO} we demonstrate that, after generalization to incorporate dislocation generation and  annihilation, already \CDDO\, provides a theoretical foundation for describing early stages of work hardening.  \CDDO\, is, however, a local plasticity theory and therefore can not describe phenomena that are explicitly related to spatial transport of dislocations. To correctly capture the spatial distribution of dislocations and the related fluxes in an inhomogeneous microstructure one needs to consider the evolution equations of \AI\, and/or of \AII. Closing the evolution equations at the level of \AI\,, or of \AII\, yields the the first order \CDDI\, and second order  \CDDII\, theories respectively. The DODF of these theories have a more complex structure that allows for directional anisotropy which we discuss in  \appref{app:CDDI} and \appref{app:CDDII} together with the derivation of the corresponding annihilation terms for directionally anisotropic dislocation arrangements. 

\section{Dynamic dislocation annihilation}

If dislocation segments of opposite orientation which belong to different dislocation loops closely approach each other, they may annihilate. This process leads to a merger of the two loops. The mechanism that determines the reaction distance is different for dislocations of near-screw and near-edge orientations: 
\begin{enumerate}
	\item Two near-screw dislocations of opposite sign,
	gliding on two parallel planes, annihilate by cross slip of one of them. 
	\item Two near-edge dislocations annihilate by spontaneous formation and disintegration of a very narrow  unstable dislocation dipole when the attractive elastic force between two dislocations exceeds the force required for dislocation climb. As opposed to screw annihilation this process generates interstitial or vacancy type point defects.
\end{enumerate}
This difference results in  different annihilation distances for screw and edge segments. The dependency of the maximum annihilation distance $y_a$ between line segments on applied stress and dislocation line orientation $\vp$ is well known \citep{ Kusov86_PSSB,Paus13_AM}. For instance, \citet{Essmann79_PMA} observed that at low temperatures (smaller than 20\% of the melting temperature), the annihilation distance changes from around  $~1.5{\rm nm}$ for pure edge dislocations to around $50{\rm nm}$ for pure screws in copper.  In CCT, dislocations of different orientation can by definition not coexist in the averaging volume, which thus is directly acting as the annihilation volume for all dislocations. Hence, it is difficult to account for differences in the annihilation behaviour of edge and screw dislocations. 

\subsection[Dislocation annihilation in CDD]{Dislocation annihilation in continuum dislocation dynamics}

\subsubsection{Straight parallel dislocations}

Coarse-grained continuum theories that allow for the coexistence of dislocations of different orientations within the same volume element require a different approach to annihilation. Traditionally this approach has used analogies with kinetic theory where two `particles' react if they meet within a reaction distance $y_a$. Models such as the one proposed by \cite{Arsenlis04_JMPS} formulate a similar approach for dislocations by focusing on encounters of straight lines which annihilate once they meet within a reaction cross-section (annihilation distance) leading to bi-molecular annihilation terms (\figref{fig:annihilation-rate}(left)). However, dislocations are not particles, and in our opinion the problem is better formulated in terms of the addition of dislocation density vectors within an 'reaction volume' that evolves as dislocations sweep along their glide planes. For didactic reasons we first consider the well-understood case of annihilation of straight parallel dislocations in these terms (\figref{fig:annihilation-rate}(left)). We consider positive dislocations of density vector $\Brho^+ = \Be_a \rho^+ $ and negative dislocations of density vector $\Brho^- = -\Be_a \rho^-$. During each time step $dt$, each positive dislocation may undergo reactions with negative dislocations contained within a differential annihilation volume $V_a = 4 y_a v s dt$ where $s$ is the dislocation length, which for straight dislocations equals the system extension in the dislocation line direction. The factor 4 stems from the fact that the annihilation cross section is $2Y_a$, and the relative velocity $2v$. The total annihilation volume in a reference volume $\Delta V$ associated with positive dislocations is obtained by multiplying this volume with the dislocation number $N^{+}$. The positive dislocation density in $\Delta V$ is $\rho^{+} = N^{+} w/\Delta V$ where $w$ is the average line segment length. Hence, the differential annihilation volume fraction (differential annihilation volume divided by reference volume) for positive dislocations is 
\begin{align}
f_a^{+}=V_a \frac{N^{+}}{\Delta V} = 2 y_a v \rho^{+}
\label{eq:fa}
\end{align}
Annihilation is now simply tantamount to replacing, within the differential annihilation volume, the instantaneous values of $\Brho^+$ and $\Brho^-$ by their vector sum. This summation reduces the densities of both positive and negative dislocations by the same amount. The average density changes in the reference volume $\Delta V$ are obtained by multiplying the densities with the respective annihilation volume fractions of the opposite 'species' and summing over positive and negative dislocations, hence
\begin{equation}
\frac{d\rho^+}{dt} = \frac{d\rho^-}{dt} = - (f^+ \rho^- + f^- \rho^+) = - 4 y_a v \rho^+ \rho^- .
\end{equation}
This result is symmetrical with respect to positive and negative dislocations.  

\subsubsection{Recombination of non-parallel dislocations}

\begin{figure}
	\centering
	\includegraphics[width=0.9\linewidth]{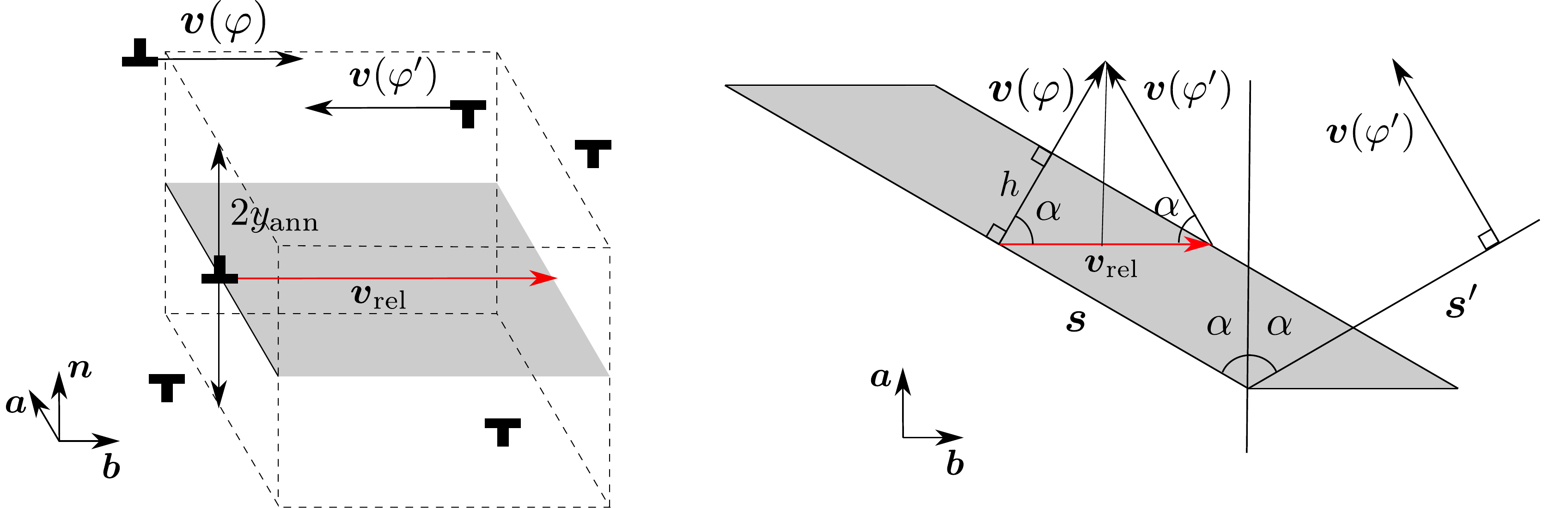}
	\caption[Annihilation of straight dislocations]{Left: Differential annihilation volume  of opposite edge dislocations is determined by multiplying their relative velocity $v_{\rm rel}=2v$ w.r.t. each other with the  line segment length $w$ and the annihilation window $2y_{\rm cb}$ and time step $\delta t$: $V_a=4 v y_{\rm ann} w d t$  . Right: Similarly, differential recombination volume of segments with orientation $\vp$ and  $\vp'=\pi+\vp-2 \alpha_{\Bl\Bl'}$ is determined by multiplying their relative velocity $v_{\rm rel}=2v \cos( \alpha_{\Bl\Bl'})$ w.r.t. the each other with the projected line segment $w \cos( \alpha_{\Bl\Bl'})$ which is perpendicular to relative velocity  and the annihilation window $2y_{\rm cb}$ and time step $\delta t$: $V_a=4 v y_{\rm ann} \cos^2( \alpha_{\Bl\Bl'}) w dt$ . }
	\label{fig:annihilation-rate}
\end{figure}

Our argument based on the differential annihilation volume can be straightforwardly generalized to families of non-parallel dislocations. 
We first consider the case where the annihilation distance does not depend on segment orientation. We consider two families of dislocation segments of equal length $s$, with directions $\Bl$ and $\Bl'$ and densities $\rho_{\Bl}$ and $\rho_{\Bl'}$. The individual segments are characterized by segment vectors $\Bs = \Bl s$ and $\Bs' = \Bl' s$ (for generic curved segments we simply make the transition to differential vectors ${\rm d}\Bs = \Bl {\rm d}s$ and ${\rm d}\Bs' = \Bl' {\rm d}s$).  The segments are moving at velocity $v$ perpendicular to their line direction (\figref{fig:annihilation-rate}(right)). 

The argument then runs in strict analogy to the previous consideration, however, since the product of the reaction is not zero we speak of a recombination rather than an annihilation reaction. Furthermore, the differential reaction (recombination) volume is governed not by the absolute velocity of the dislocations but by the velocity at which either of the families sweeps over the other. This relative velocity is given by $v_{\rm rel} = 2 v \cos\alpha_{\Bl\Bl'}$ where $2 \alpha_{\Bl\Bl'} = \pi - \psi$ and $\psi$ is the angle between the velocity vectors of both families (\figref{fig:annihilation-rate}(right)). The recombination area that each segment sweeps by its relative motion to the other segment is thus given by  $A_a = 2 v \cos\alpha_{\Bl\Bl'} s dt$. The differential recombination volume is then in analogy to Eq. \ref{eq:fa} given by
\begin{align}
f_r^{\Bl} = 4 y_a v \cos^2 (\alpha_{\Bl\Bl'}) \rho_{\Bl}.
\label{eq:fr}
\end{align} 
Within this volume fraction we identify for each segment of direction $\Bl$ a segment of orientation $\Bl'$ of equal length $s$ and {\em replace} the two segments by their vector sum (in the previously considered case of opposite segment directions, this sum is zero). Hence, we reduce, within the differential recombination volume, both densities by equal amounts and add new segments of orientation $\Bl''$and density $\rho_{\Bl''} s''$ where $\Bl''$ and $s''$ fulfil the relations:
\begin{align}
\Bs''=\Bs+\Bs',\, s'' = |\Bs''|,\, \Bl'' = \frac{\Bs''}{s''}
\end{align}
We can now write out the rates of dislocation density change due to recombination as
\begin{align}
\frac{d\rho_{\Bl}}{dt} = \frac{d\rho_{\Bl'}}{dt} = - 4 y_a v \rho_{\Bl} \rho_{\Bl'} \cos^2(\alpha_{\Bl\Bl'}),\nonumber\\
\frac{d\rho_{\Bl''}}{dt} = 4 y_a v \rho_{\Bl} \rho_{\Bl'} \cos^2(\alpha_{\Bl\Bl'}) s''(\Bl,\Bl').
\label{eq:recombine_l}
\end{align}
For dislocations of opposite line directions, $\alpha_{\Bl\Bl} = 0$ and $s''=0$, hence, we recover the previous expression for annihilation of parallel straight dislocations. For dislocations of the same line direction, $\alpha_{\Bl\Bl'}=\pi/2$, $s'' = 2s$, and $\Bl'' = \Bl' = \Bl$, hence, there is no change in the dislocation densities. 

\subsubsection{Recombination of loops triggered by cross slip}
\label{subsec:coss_slip_annihilation}
\begin{figure}
	\centering
	\includegraphics[width=0.7\linewidth]{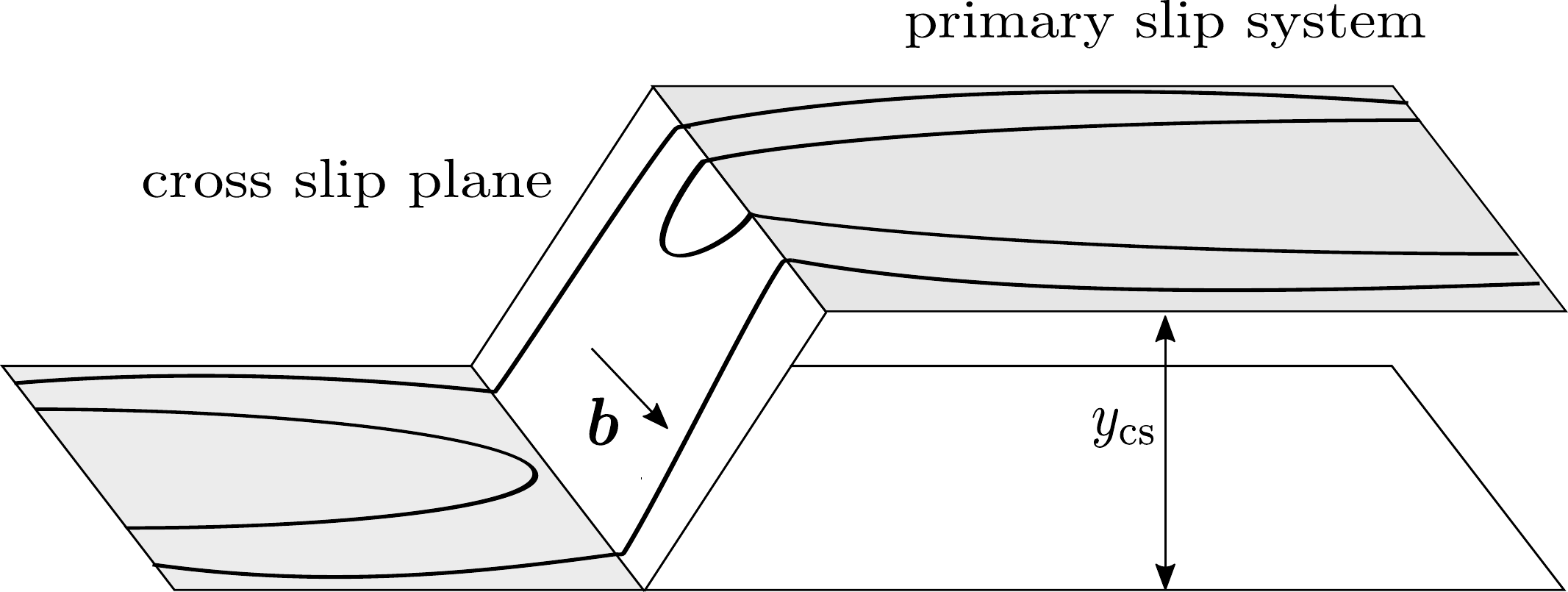}
	\caption[Two dislocation loops in a cross slip configuration]{Dislocation loops in a cross slip configuration. After cross slip annihilation two semi-loops are connected by collinear jogs moving in Burgers vector direction. }
	\label{fig:cross-slip}
\end{figure}
\begin{figure}
	\centering
	\includegraphics[width=0.7\linewidth]{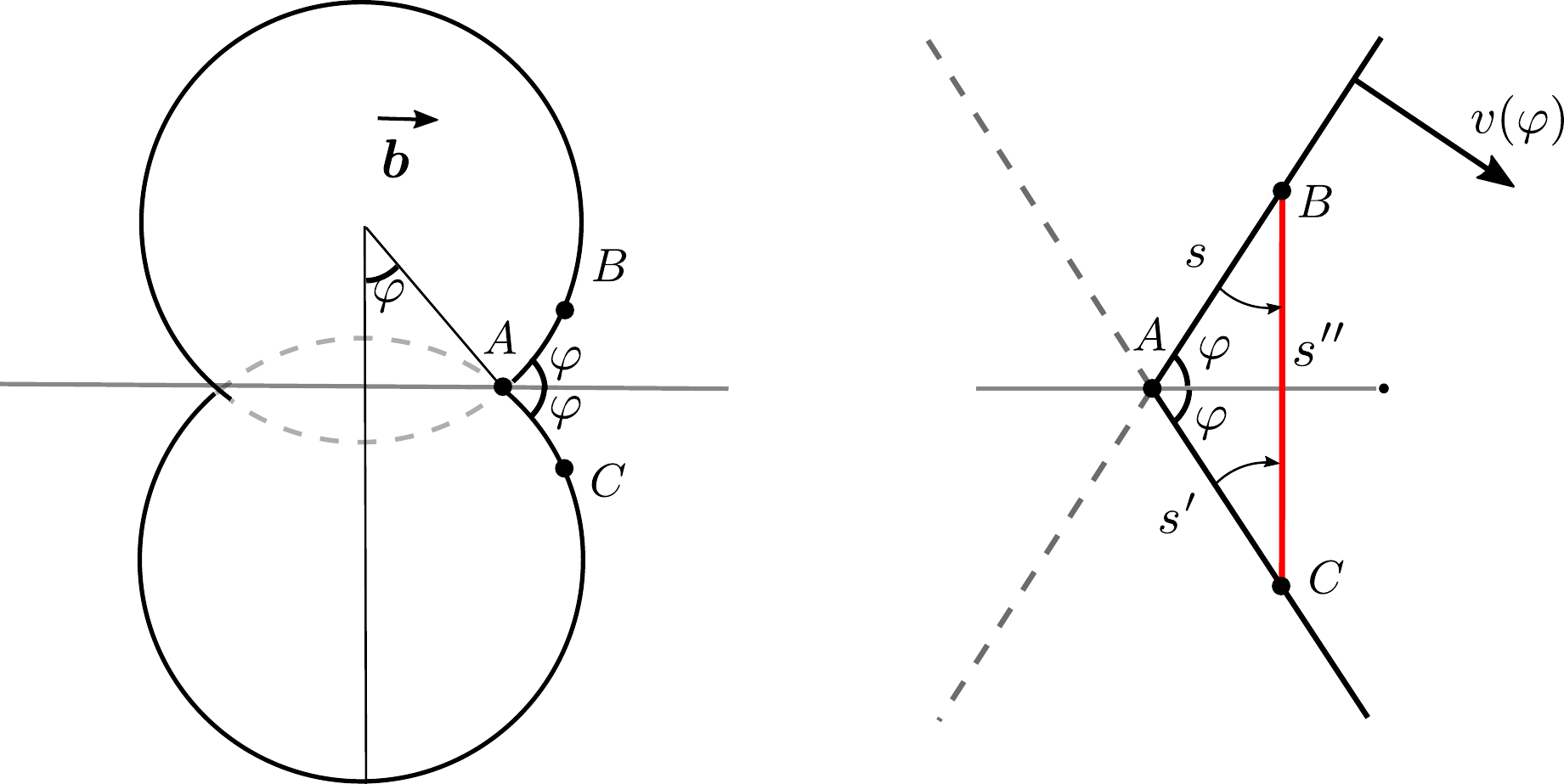}
	\caption[Top view of a cross slip induced recombination process]{	Top view of a cross slip induced recombination process. Left: cross slip initiates the  annihilation of near-screw  segments of two merging dislocation loops.  The dashed lines shows the annihilated parts of the loops. After the initiation of the cross slip, loops continue to merge by interaction between segments AB($\Bs(\vp)$) and AC ($\Bs(\vp')$). Right:  Recombination of segments  $\Bs(\vp)$ and $\Bs(\vp')$ generates a new segment $\Bs(\vp'')$ with edge orientation which changes the total dislocation density and mean orientation.}
	\label{fig:annihilation_loop}
\end{figure}
We now generalize our considerations to general non-straight dislocations, i.e., to ensembles of loops. We first observe that the relations for straight non-parallel dislocations hold locally also for curved dislocations, provided that the dislocation lines do not have sharp corners. For curved dislocations we characterize the dislocation ensemble in terms of its DODF of orientation angles, i.e., we write 
\begin{align}
&\Bl = \Bl(\vp) = (\cos\vp,\sin\vp) \quad,\quad \rho_{\Bl} = \rho p(\vp);\nonumber\\
&\Bl' = \Bl(\vp') = (\cos\vp',\sin\vp') \quad,\quad , \rho_{\Bl'} = \rho p(\vp');\nonumber\\
&\Bl'' = \Bl(\vp'') = (\cos\vp'',\sin\vp'') \quad,\quad , \rho_{\Bl''} = \rho p(\vp'');\nonumber\\
&\alpha_{\Bl\Bl'}=\alpha(\vp,\vp').
\end{align}
We now first consider the recombination of loops initiated by cross slip of screw dislocation segments. This process is of particular importance because the annihilation distance $y_{\rm cs}$ for near-screw dislocations is almost two orders of magnitude larger than for 
other orientations \citep{Paus13_AM}. The recombination process is initiated if two near-screw segments which are oriented within a small angle $\vp_{\rm a} \in [- \Delta \vp, \Delta \vp]$ from the screw orientations $\vp = 0$ and $\vp=\pi$ pass within the distance $y_{\rm cs}$ (\figref{fig:cross-slip}). Mutual interactions then cause one of the near-screw segments to cross slip and move on the cross-slip plane until it annihilates with the other segment. However, it would be erroneous to think that cross slip only affects the balance 
of near-screw oriented segments: Cross slip annihilation of screw segments connects two loops by a pair of segments which continue to move in the cross-slip plane. We can visualize the geometry of this process  by considering the projection of the resulting configuration on the primary slip plane.  \figref{fig:annihilation_loop}(left) depicts the top view of a situation some time after near-screw segments of two loops moving on parallel slip planes have merged by cross slip. As the loops merge, the intersection point $A$ -- which corresponds to a collinear jog  in the cross slip plane that connects segments of direction $\Bl(\vp)$ and $\Bl(\vp')$ in the primary slip planes -- moves in the Burgers vector direction. Hence, the initial cross slip triggers an ongoing recombination of segments of both loops as the loops continue to expand in the primary slip system \citep{Devincre07_SM}. 

We note that the connecting segments produce slip in the cross-slip plane. The amount of this slip can be estimated by considering a situation well after the recombination event, when the resulting loop has approximately spherical shape with radius $R$. The slipped area in the primary slip plane is then $\pi R^2$, and the area in the cross slip plane is, on average, $R y_{\rm cs}/2$. Hence, the ratio of the slip amount in the primary and the cross slip plane is of the order of $2 \pi R/y_{\rm cs} \approx 2 \pi \rho/(y_{\rm cs}q)$. We will show later in Section 5 that, for typical hardening processes, the amount of slip in the cross slip plane caused by recombination processes can be safely neglected.

Comparing with Figure \ref{fig:annihilation-rate} we see that, in case of cross slip induced recombination, the two recombining segments fulfil the orientation relationship $\vp' = \pi -\vp$ and that the angle $\alpha_{\Bl\Bl'}$ and the length of the recombined segment are given by 
\begin{align}
&\alpha_{\Bl\Bl'}= \vp,\\
&\vp'' = \frac{\pi}{2} {\rm sign}(\pi-\vp), \\
&s'' = |\Bl + \Bl'| = 2|\sin\vp|.
\end{align}
We now make an important conceptual step by observing that, if two segments pertaining to different loops in the configuration shown in Figure \ref{fig:annihilation_loop} are found at distance less than $y_{\rm cs}$, then a screw annihilation event must have taken place in the past. Hence, we can infer from the current configuration that in this case the loops are recombining. The rates for the process follow from \eqref{eq:recombine_l} as
\begin{align}
\frac{d\rho(\vp)}{dt} = \frac{d\rho(\vp')}{dt} = - 4 y_{\rm cs} v \rho^2 p(\vp)p(\vp') \cos^2(\vp),\nonumber\\
\frac{d\rho(\vp'')}{dt} = 8 y_{\rm cs} v \rho^2 p(\vp)p(\vp')  \cos^2(\vp) |\sin\vp|.
\label{eq:recombine__phiscrew}
\end{align}
Multiplying \eqref{eq:recombine__phiscrew} with the appropriate power tensors of the line orientation vectors and integrating over the orientation window where cross slip is possible gives the change of alignment tensors due to cross slip induced recombination processes: 
\begin{align}\label{eq:Ann_rhok_cross}
\partial_t \rho^{(k)}_{\rm cs} &= -4 y_{\rm cs} v \rho^2 \oint \oint \Theta(\Delta \vp - |\vp + \vp' - \pi|) 
\cos^2(\vp) \left[\Bl^{(k)}(\vp) - \left|\sin(\vp)\right| \Bl^{(k)}(\pi/2)\right]  {\rm d}\vp'{\rm d}\vp\nonumber\\
 &-4 y_{\rm cs} v \rho^2 \oint \oint \Theta(\Delta \vp - |\vp + \vp' - 3\pi|) 
\cos^2(\vp) \left[\Bl^{(k)}(\vp) - \left|\sin(\vp)\right| \Bl^{(k)}(3\pi/2)\right]  {\rm d}\vp'{\rm d}\vp.
\end{align} 
Here $\Theta$ is Heaviside's unit step function that equals 1 if its argument is positive or zero, and zero otherwise. Hence, 
$\vp'$ must be located within $\Delta \vp$ from $\pi - \vp$ if $\vp$ is less than $\pi$, and within $\Delta \vp$ from $3 \pi - \vp$ 
if $\vp$ is bigger than $\pi$. 

\subsubsection{Isotropic recombination of general dislocations by climb}
\begin{figure}[t]
	\centering
	\includegraphics[width=0.7\linewidth]{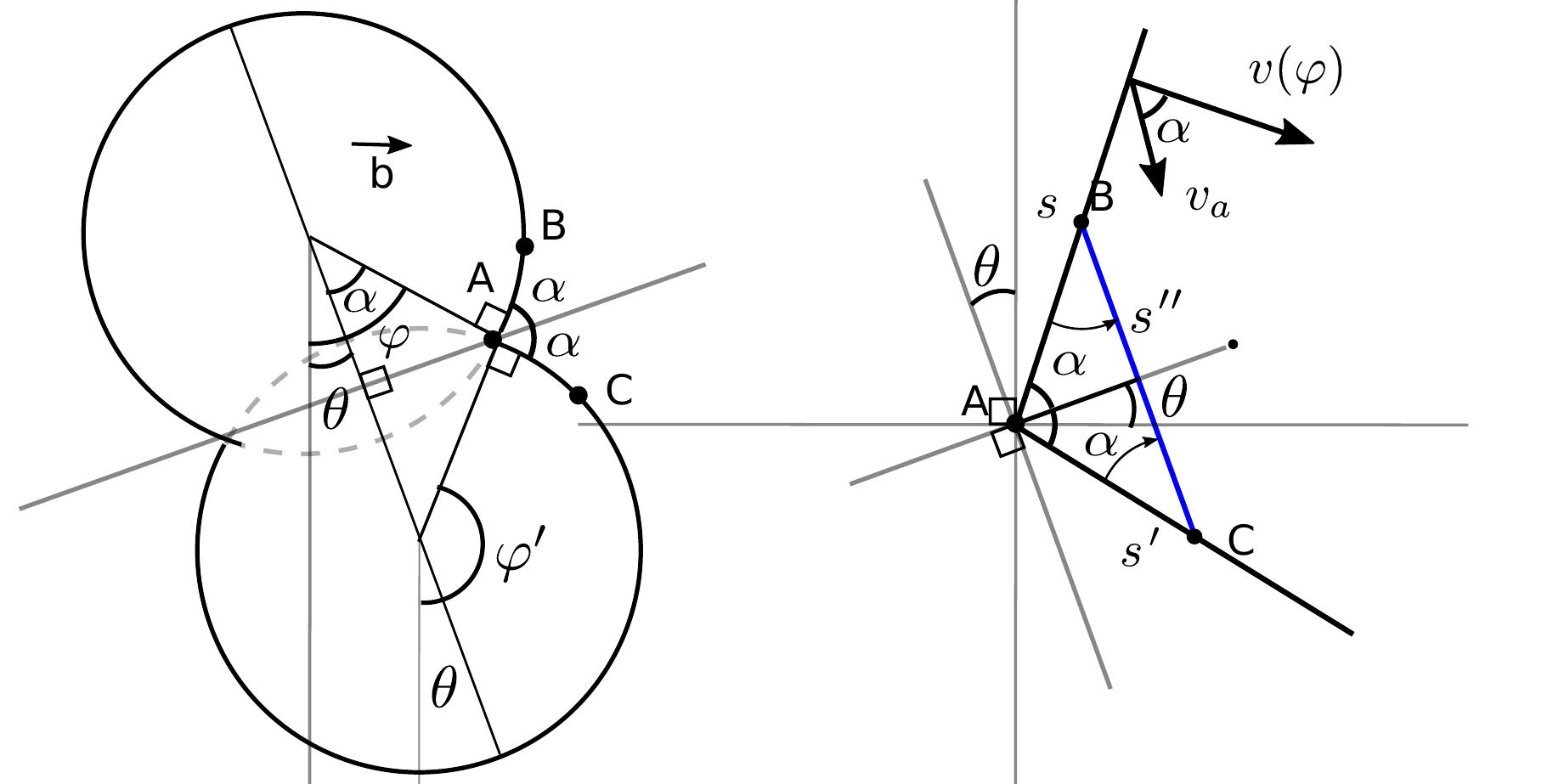}
	\caption[Top view of a climb slip annihilation process]{Left: Two dislocation loops are merging by climb annihilation initiated at segments with angles $\theta$ and $\pi+\theta$. Right:  Interaction between segments  $s(\vp)$ and $s(\vp'=\pi+2\theta-\vp)$ generates a news segment $s(\vp'')$ with  orientation perpendicular to $\theta$.}
	\label{fig:annihilation_all_degrees}
\end{figure}

Next we consider recombination by climb  which we suppose to be possible for dislocations of any orientation that are within a direction-independent cross-section $2 y_{\rm cb}$ of others. Hence, the process is - unlike cross slip - isotropic in the sense that an initially isotropic orientation distribution will remain so, and recombination can occur between segments of any orientation provided they find themselves within a distance of less than $y_{\rm cb}$. To analyse this process, we focus on the plane of symmetry that bisects the angle between both segments. This plane is at an angle $\theta$ from the screw dislocation orientation, see \figref{fig:annihilation_all_degrees}. 
Now, if we rotate the picture by $-\theta$, it is clear that the geometry of the process is exactly the same as in case of cross slip induced 
recombination, and that only the appropriate substitutions need to be made. The following geometrical relations hold:
\begin{align}
\theta(\vp,\vp') &= \frac{\vp'+\vp-\pi}{2},\\
\alpha(\vp,\vp') &= \frac{\vp-\vp'+\pi}{2},\\
\vp''&=\frac{\vp+\vp'}{2}, \\
s'' &= 2\left|\cos\left(\frac{\vp-\vp'}{2}\right)\right|.
\end{align}
According to \eqref{eq:recombine_l} the rate of recombination between segments of directions $\vp$ and $\vp'$ then leads to the following density changes:
\begin{align}
\frac{d\rho(\vp)}{dt} = \frac{d\rho(\vp')}{dt} = - 4 y_a v \rho^2 p(\vp) p(\vp') \sin^2 \left(\frac{\vp-\vp'}{2}\right),\nonumber\\
\frac{d\rho(\vp'')}{dt} = 8 y_a v \rho^2 p(\vp)p(\vp') \sin^2\left(\frac{\vp-\vp'}{2}\right) \left|\cos\left(\frac{\vp-\vp'}{2}\right)\right|.
\label{eq:recombine_phi}
\end{align}
Multiplying \eqref{eq:recombine_phi} with the appropriate power tensors of the line orientation vectors and integrating over all orientations gives the change of alignment tensors due to climb recombination processes: 
\begin{align}
\partial_t \Brho^{(k)} = - 4 y_{\rm cb} v \rho^2 \oint \oint p(\vp)p(\vp') 
\sin^2\left(\frac{\vp-\vp'}{2}\right)\left[\Bl^{(k)}(\vp) -\left|\cos\left(\frac{\vp-\vp'}{2}\right)\right| \Bl^{(k)}\left(\frac{\vp+\vp'}{2}\right)\right]  {\rm d}\vp'{\rm d}\vp
\label{eq:Ann_DODF_all_rec}
\end{align}
In particular, the rates of change of the lowest-order tensors are
\begin{align}
\partial_t \rho &= - 4 y_{\rm cb} v \rho^2 \oint \oint p(\vp)p(\vp') 
\sin^2\left(\frac{\vp-\vp'}{2}\right)\left[1 -\left|\cos\left(\frac{\vp-\vp'}{2}\right)\right|\right]  {\rm d}\vp'{\rm d}\vp\\
\partial_t \Brho &= 0.
\label{eq:Ann_DODF_01}
\end{align}
The latter identity is immediately evident if one remembers that $\Brho$ is the vector sum of all dislocation density vectors in a volume, hence, it cannot change if any two of these are added up and replaced with their sum vector. 

\section{Dynamic dislocation sources}
During early stages of plastic deformation of a well-annealed crystal ($\rhot\approx10^6[\rm{m}^{-2}]$), the dislocation density can increase by several orders of magnitude. This increase of dislocation density contributes to many different phenomena such as work hardening. Therefore, no dislocation theory is complete without adequate consideration of the multiplication problem. In CDD, multiplication in the sense of line length increase by loop expansion occurs automatically because the kinematics of curved lines requires so, however, the generation of new loops is not accounted for, which leads to an incorrect hardening kinetics. In this section, we discuss several  dynamic mechanisms that increase the loop  density by generating new dislocation loops. 

First we introduce the well-known Frank-Read source and how we formulate it in a continuous sense in the CDD framework. Frank-Read sources are fundamental parts of the cross-slip and glissile junction multiplication mechanisms which play an important role in work hardening. Therefore we use the Frank-Read source analogy to discuss the kinematic aspects of these mechanisms and the necessary steps for incorporating them into the CDD theory. We first note that the Mura equation, if applied to a FR source configuration with sufficiently high spatial resolution to a FR source, captures the source operation naturally without any further assumptions, as shown by the group of Acharya \citep{Varadhan06_MSMSE}.
Like the problem of annihilation, the problem of sources arises in averaged theories where the spatial structure of a source can not be resolved. To overcome this problem, \citet{Hochrainer06_Phd} proposed a formulation for a continuous FR source distribution in the context of the higher-dimensional CDD. \citet{Sandfeld11_ACP} described the operation of a single FR source in the context of lowest-order CDD theory as a discrete sequence of loop nucleation events.  \citet{Acharya01_JMPS} generalizes CCT to add a source term into the Mura equation. This term might represent the nucleation of dislocation loops of finite area {\em ex nihil} which can happen at stresses close to the theoretical shear strength, or through diffusion processes which occur on relatively long time scales and lead to prismatic loops \citep{Messerschmidt03_MCP,Li15_NM}. Neither process is relevant for the normal hardening behavior of metals. 

\subsection[Frank-Read source]{Frank-Read sources}
\begin{figure}[t]
	\centering
	\includegraphics[width=.4\linewidth]{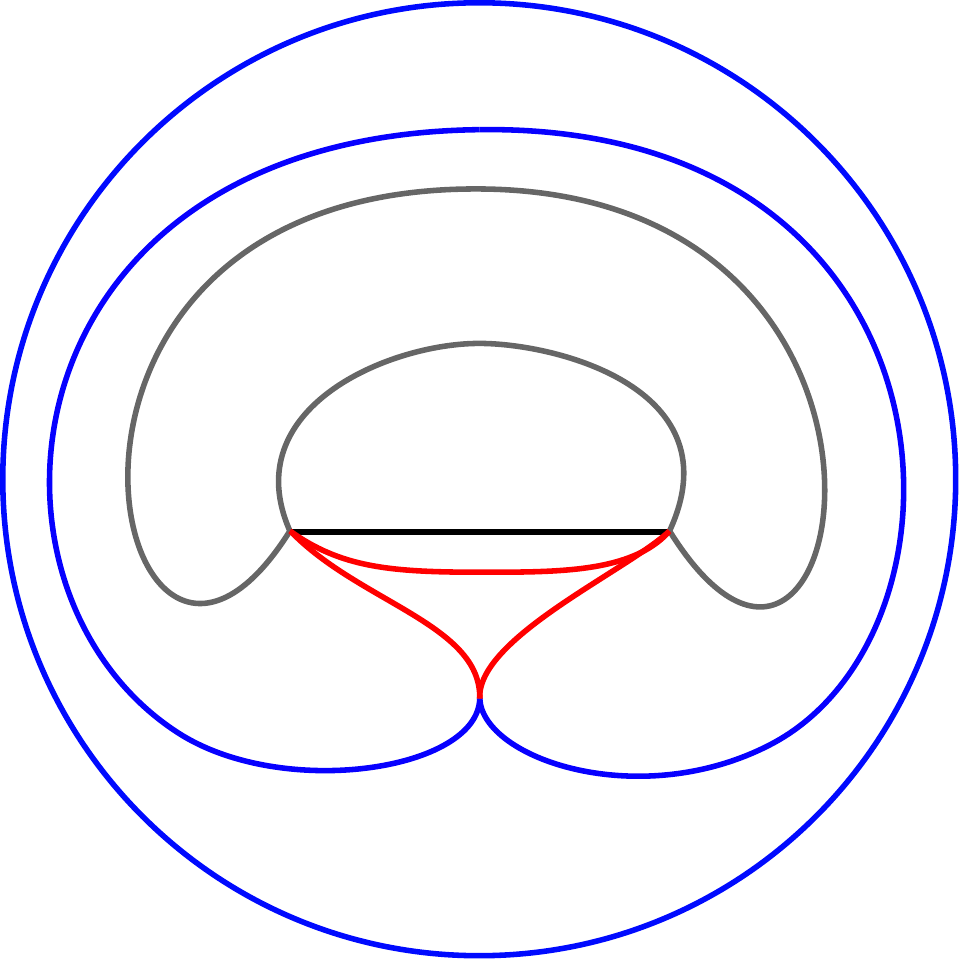}	
	\caption[Frank-Read source]{Activation of a Frank-Read source: A dislocation segment (black) pinned at both ends bows out under applied stress and creates a metastable half-loop. If the  shear stress acting on the source is higher than a critical value, this semi-loop expands further and rotates around the pinned ends. The recombination of loop segments (red) then generates a new complete loop (blue) and restores the original configuration.}
	\label{fig:FR-source} 
\end{figure}  
The  main  mechanism  for generation of new dislocation loops in low stress conditions was first suggested by  \citet{Frank50_PR}. Here we propose a phenomenological approach to incorporate this mechanism into CDD. A  Frank-Read source is a dislocation segment with pinned end points, e.g. by interactions with other defects or by changing  to a slip plane  where  it  is  not  mobile. Under stresses higher than a critical stress, the  segment bows out and generates a new dislocation loop and a pinned segment identical to the initial segment. Therefore, a Frank-Read source can  successively generate closed dislocation loops \figref{fig:FR-source}. A Frank-Read source can only emit dislocations when the shear stress is higher than a critical stress needed to overcome the maximum line tension force \citep{Hirth82_Book}:
\begin{align}
\label{eq:source_critical_shear_stress}\sigma_{\rm cr}\approx\frac{G b}{r_{\rm FR}},
\end{align}
where the radius of the metastable loop is half the source length, $r_{\rm FR}=\frac{L}{2}$.
The activation rate of Frank-Read sources has been subject of several studies. \citet{Steif79_MSE} found that in typical FCC metals, the multiplication process is controlled by the activation rate at the source, where the net driving force is minimum due to high line tension. 
The nucleation time can be expressed in a universal plot of dimensionless stress $\sigma^*=\sigma L/G b$ vs dimensionless time $t^*=t_{\rm nuc}\sigma b/B L$, where $t_{\rm nuc}$ is the nucleation time and $B$ is the dislocation viscous drag coefficient. For a typical $\sigma^*\approx4$, the reduced time becomes $t^*\approx10$ \citep{Hirth82_Book}. However, this exercise may be somewhat pointless because the stress at the source cannot be controlled from outside, rather, it is strongly influenced by local dislocation-dislocation correlations, such as the back stress from previously emitted loops. Such correlations have actually a self-regulating effect: If the velocity of dislocation motion near the source for some reason exceeds the velocity far away from the source, then the source will emit dislocations rapidly which pile up close to it and exert a back stress that shuts down the source. Conversely, if the velocity at the source is reduced, then previously emitted dislocations are convected away and the back stress decreases, such that source operation accelerates. The bottom line is, the source will synchronize its activation rate with the motions of dislocations at a distance. In our kinematic framework which averages over volumes containing many dislocations, it is thus reasonable to express the activation time in terms of the average dislocation velocity $v=\sigma b /B$ as:
\begin{align}
\label{eq:FR_activation_rate}
\tau=\eta r_{\rm FR}/v.
\end{align}
\eqref{eq:FR_activation_rate} implies that the activation time is equal to the time that an average dislocation takes to travel $\eta$ times the Frank-Read source radius before a new loop can be emitted. In discrete dislocation dynamics (DDD) simulations, a common practice for creating the initial dislocation structure  is to consider a fixed number of grown-in Frank-Read sources distributed over the different slip systems  \citep{Motz09_AM}. Assuming that the length of these sources is $2 r_{\rm FR}$ and the density of the source dislocations is $\rho_{\rm FR}$, then their volume density is $n_{\rm FR} = \rho_{\rm FR}/(2 r_{\rm FR})$. The activation rate is given by the inverse of the nucleation time: 
\begin{align}
\label{eq:FR_rate}
\nu_{{\rm FR}}&=  v / (\eta r_{\rm FR}).
\end{align}
The operation of Frank-Read sources of volume density $n_{\rm FR}$ increases the curvature density by $2 \pi$ 
times the loop emission rate per unit volume, hence
\begin{align}
\label{eq:qt_FR_source_rate}
\dot{q}_{\rm fr}&= 2\pi n_{\rm FR} \nu_{\rm FR} = \pi v\frac{\rho_{\rm FR}}{\eta r_{\rm FR}^2}.
\end{align}
We note that no corresponding terms enter the slip rates, or the evolution of the alignment tensors, which are fully described
by terms characterizing motion of already generated dislocations.

Source activity has important consequences for work hardening. The newly created loops have high curvature of the order of the inverse loop radius, hence they are more efficient in creating line length than old loops that have been expanding for a long time. This effect of increasing the average curvature of the dislocation microstructure is of major importance for the work hardening kinetics.

\subsection{Double-cross-slip sources}

\citet{Koehler52_PR} suggested the double-cross-slip mechanism as a similar mechanism to a Frank-Read source that can also repeatedly emit  dislocation loops.   In  double-cross-slip,   a  screw  segment that is gliding on the plane with maximum resolved shear stress (MRSS) and is blocked by an obstacle  cross-slips to a slip plane with lower MRSS. After passing the obstacle it cross slips back to the original slip system and produces two super jogs connecting the dislocation lines. These two super-jogs may act as pinning points for the dislocation and in practice produce Frank-Read like sources. The double-cross-slip source is the result of the interaction of dislocations on different slip planes and therefore a dynamic process. 

Several DDD studies such as \cite{Hussein15_AM} have tried to link the number of double-cross-slip sources in the bulk and on the boundary of grains to the total dislocation density. They observed that the number of double-cross-slip sources increases with dislocation  density and  specimen size. However, these studies fall short of identifying an exact relation between the activation rate of cross-slip-sources and system parameters.
\begin{figure}[t]
	\centering
	\label{fig:CS_source}
	\includegraphics[width=.7\linewidth]{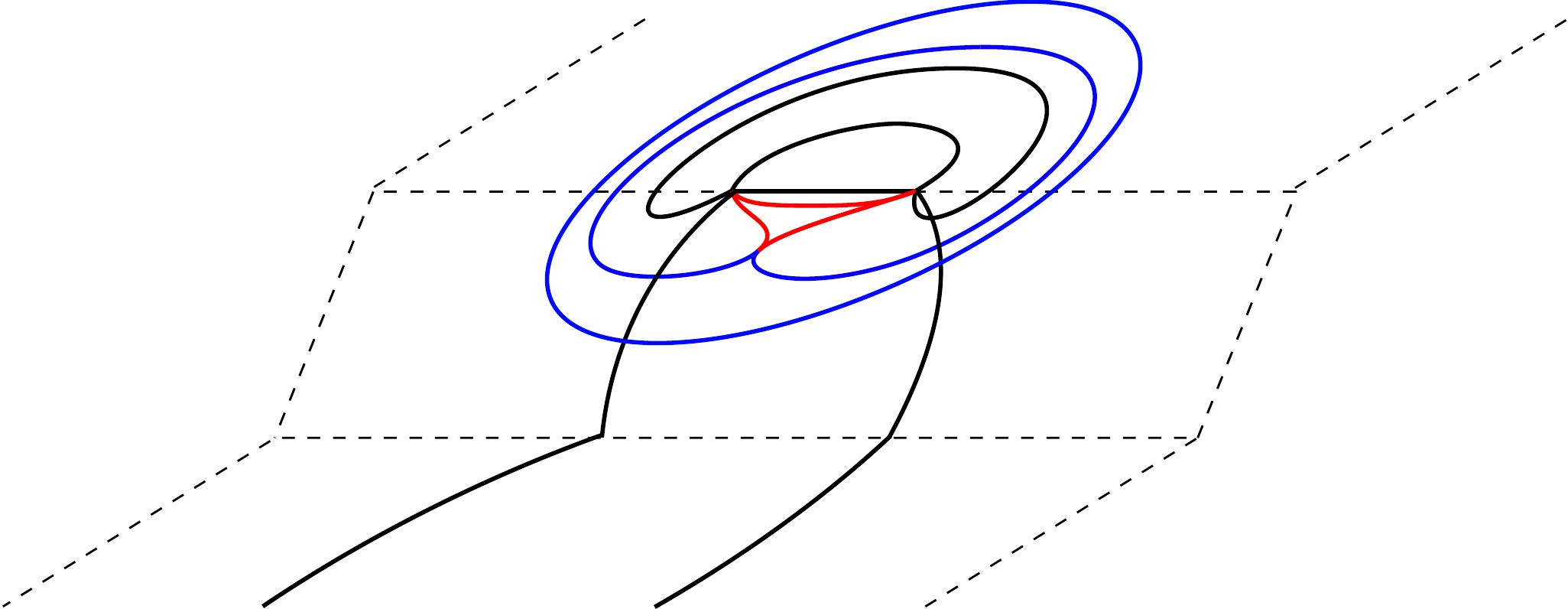}
	\caption[Cross-Slip source]{A double cross-slipped segment may act as a Frank-Read source on a parallel slip plane.} 
\end{figure}  
In the following we introduce a model for incorporating this process into CDD. The  density of screw dislocations $\rho_{\rm s}$ on a slip system is:
\begin{align}
	\rho_{\rm s}=\int_{-\Delta \varphi}^{\Delta \varphi} \rho(\vp) + \int_{\pi-\Delta \varphi}^{\pi+\Delta \varphi} \rho(\vp),\label{eq:intscrew}
 \end{align} which in general is a function of dislocation moments functions. For the case of isotropic DODF this can be simplified to $\rho_{\rm s} = 4 \Delta \vp (\rho(\vp     =0) + \rho(\vp     =\pi))= \frac{4\Delta\vp}{2\pi}\rho$. We assume that a fraction $f_{\rm dcs}$ of this density is in the form of double-cross-slipped and pinned segments. Hence, the source density is $n_{\rm dcs} = \rho_{\rm s}/r_{\rm dcs}$, where the pinning length of the cross slipped segments is of the order of the dislocation spacing, $r_{\rm dcs} = 1/\sqrt{\rhotot}$ with $\rhotot    = \sum_{\varsigma} \rho$. Otherwise we assume for the cross-slip source exactly the same relations as for the grown-in sources of density $\rho_{\rm FR}$ and radius $r_{\rm FR}$. Thus, the generation rate of curvature density becomes:
\begin{align}
\label{eq:qt_source_rate}
\dot q_{\rm dcs}&\approx \pi\frac{f_{\rm dcs}}{\eta} v \rho_{\rm s} \rhotot.
\end{align}

The non-dimensional numbers $f_{\rm dcs}$ and ${\eta}$ can be determined by fitting CDD data to an ensemble average of DDD simulations, or to work hardening data. While in bulk systems these parameters only depend on the crystal structure and possibly on the distribution of dislocations over the slip systems, for small systems, $f_{\rm dcs}$ and ${\eta}$ are expected to be functions of $\sqrt{\rhotot} l_{\rm s}$,  the system size (e.g. grain size) $l_{\rm s}$ in terms of dislocations spacing, because the source process may be modified e.g. by image interactions at the surface.  

\subsection{Glissile junctions}
\begin{figure}
	\centering
	\includegraphics[width=0.5\linewidth]{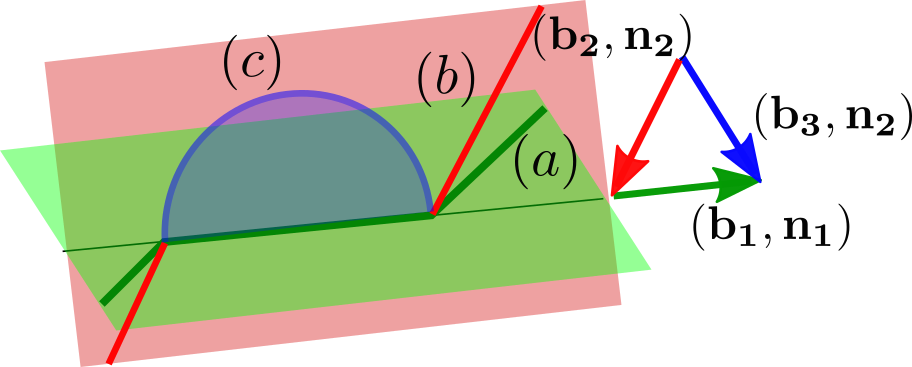}
	\caption[Glissile junction]{Glissile junction reproduced from \citet{Stricker15_AM}: Two dislocations on slip systems ($\Bb_1,\Bn_1$) and ($\Bb_2,\Bn_2$) interact and form a glissile junction acting like a Frank-Read source on slip system ($\Bb_3,\Bn_2$).} 
	\label{fig:glissilejunction}
\end{figure}

When two dislocations gliding on different slip systems ($\varsigma',\varsigma''$) intersect, it can be energetically favourable for them to  react and form a third segment called junction. Depending on the Burgers vectors and slip planes of the interacting segments this junction can be glissile (mobile) or  sessile (immobile). \figref{fig:glissilejunction} depicts the formation of a  glissile junction . The segment (a) on the slip system ($\Bb_1,\Bn_1$) interacts with the  segment (b) on the slip system ($\Bb_2,\Bn_2$) and together they produce the junction (c) on the  slip system  ($\Bb_3=\Bb_1+\Bb_2,\Bn_2$) which lies on the same glide plane as the segment (b). This mechanism produces a segment on the slip system $(\Bb_3,\Bn_2)$ with endpoints that can move and adjust the critical stress to the applied shear stress.  
Recently \cite{Stricker15_AM} studied the role of glissile junctions in plastic deformation.  They found by considering  different dislocation densities, sizes and crystal orientations of samples, that glissile junctions are one of the major contributors to the total dislocation density and plastic deformation. The action of glissile junctions can be envisaged in a similar manner as the action of cross slip sources, however, we need to take into account that only a very limited number of reactions can produce a glissile junction. Suppose that two dislocations of slip systems $\varsigma'$ and $\varsigma''$ produce a glissile junction that can act as a source on system $\varsigma$.  The density of segments on $\varsigma'$ that form junctions with $\varsigma''$ is $f_{\rm gj} \rho^{\varsigma'}\rho^{\varsigma''}/{\rhotot}$ and the length of the junctions is of the order of the dislocation spacing, $r_{\rm gj} = 1/\sqrt{\rhotot}$.  Hence we get 
\begin{align}
\label{eq:qt__cj_source_rate}
\dot q^\varsigma_{\rm gj}&\approx\sum_{\varsigma'}\sum_{\varsigma''}\pi f^{{\varsigma'}{\varsigma''}}_{\rm gj} v^\varsigma  \frac{\rho^{\varsigma'}\rho^{\varsigma''}}{\eta}.
\end{align}
We finally note that the action of dynamic sources  and recombination processes is kinematically irreversible. Consequently, by reversing the direction of the velocity, recombination mechanisms do not act as sources and vice versa. 

\section{\CDDO: a model for early stages of work hardening}
\label{sec:CDDO}
We now use the previous considerations to establish a model for the early stages of work hardening. In doing so we make the simplifying assumption that the 'composition' of the dislocation arrangement, i.e. the distribution of dislocations over the different slip systems, does not change in the course of work hardening. This is essentially correct for deformation in high-symmetry orientations but not for deformation in single slip conditions. We thus focus on one representative slip system only and assume that all other densities scale in proportion.

\begin{figure}[h]
	\includegraphics[width=0.99\linewidth]{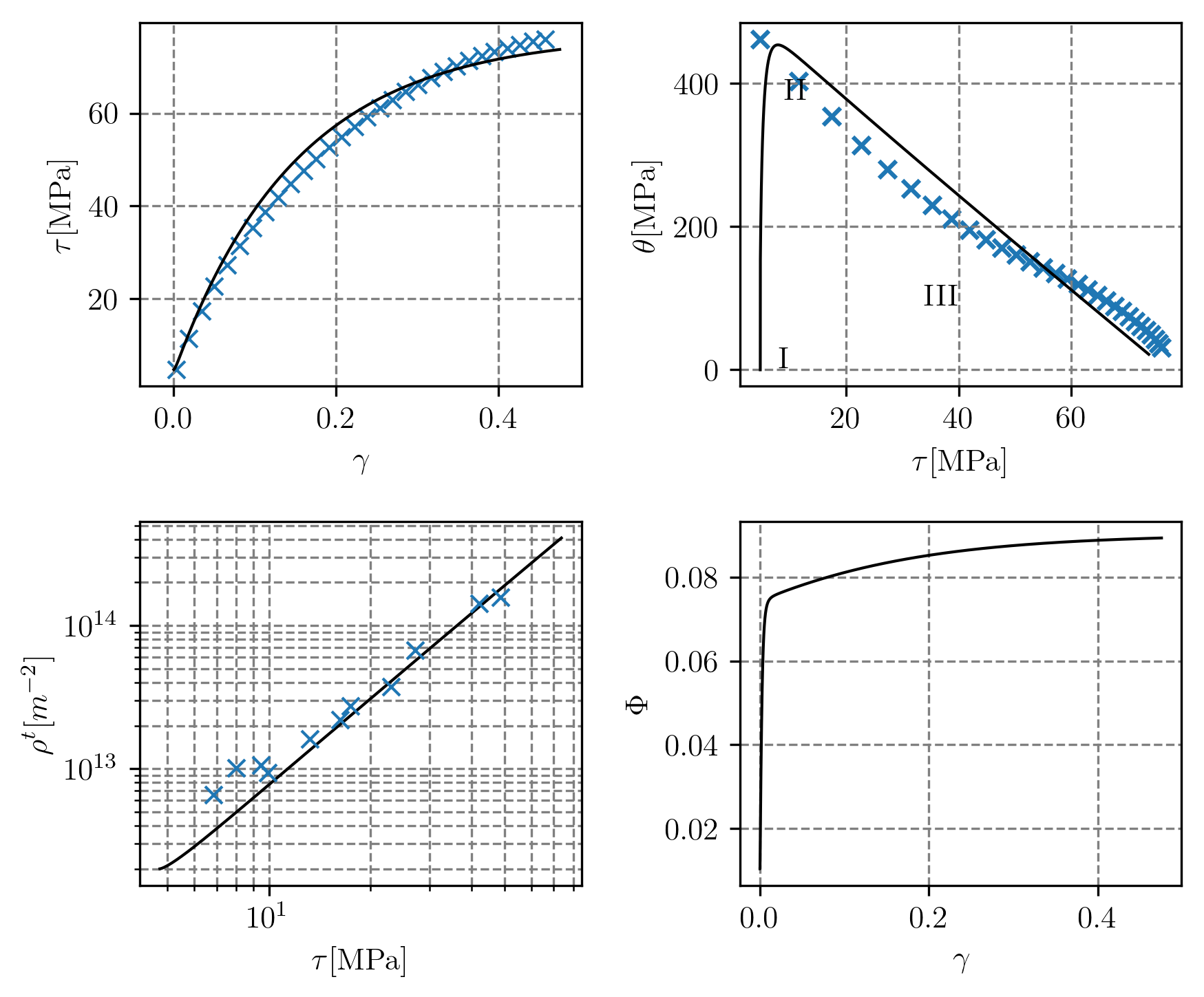}
	\caption{ First 3 stages of work hardening in cooper rolling.  Experimental measures marked by $[\times]$ obtained from \citet{Gottler73_PM}. Top-left: resolved shear stress(RSS) against plastic slip. Top-right: hardening-rate vs RSS; Hardening rate of experimental measures are obtained by fitting a 6th-order polynomial to stress-strain curve. Bottom-left: log-log plot of  dislocation density vs RSS; This plot shows that dislocation density eventually saturates as the RSS can not increase any more. Bottom-right: Dislocation-entanglement ($\Phi=q/\rho^{1.5}$) vs plastic slip.   }

	\label{fig:hardening}
\end{figure}

Since the DODF of \CDDO\, is uniform ($\rho(\vp)=\frac{\rho}{2\pi}$), the climb and cross slip recombination rates can be combined into one set of equations:
\begin{align}
\dot\rho_{\rm ann}=- 4 d_{\rm ann} v  \rho^2
\dot q_{\rm ann}= \frac{q}{\rho} \dot\rho_{\rm ann}
\end{align}
where $d_{\rm ann}$ is an effective annihilation distance. Although in DDD simulations, artificial Frank-Read sources are often used to populate a dislocation system in early stages, we consider samples with sufficient initial dislocation density where network sources (glissile junctions) are expected to dominate dislocation multiplication. Therefore, we only consider glissile junctions in conjunction with loop generation by double-cross-slip which leads to terms of the same structure. Their contribution can be combined into one equation:
\begin{align}
\dot q_{\rm src}  = \frac{c_{\rm src}}{\eta} v  \rho^2.
\end{align}
Closing the kinematic equations \eqref{eq:dA0dt} and \eqref{eq:dqtdt} at zeroth order  together with the contribution of annihilation and sources gives the semi-phenomenological \CDDO evolution equations: 
\begin{align}
\partial_t\rho &= q v + \dot\rho_{\rm ann}\nonumber\\
\partial_t q&= \dot q_{\rm ann} + \dot\qt_{\rm src}  \nonumber\\
\partial_t \gamma &= \rho b v 
\label{eq:CDDO}
\end{align}
For quasi-static loading the sum of internal stresses should balance the applied resolved shear stress. For homogeneous dislocation microstructure the dominant  internal stresses are a friction like flow stress  $\tau_{\rm f}\approx\alpha b G \sqrt{\rho}$ and a self interaction stress  associated to line tension of curved dislocations approximated as $\tau_{\rm lt}\approx T G b \frac{q}{\rho}$ where $G$ denotes the shear modulus and $\alpha$ and $T$ are non dimensional parameters \citep{Zaiser07_PM}. Therefore the applied stress becomes:
\begin{align}
	\tau_{\rm ext}&= \tau_{\rm f} +  \tau_{\rm lt}=\alpha b G \sqrt{\rho}+ T G b \frac{q}{\rho}
\end{align}
Using these relations we can build a semi-phenomenological model for work hardening. We fit the parameters of the model to the stage III hardening rate ($\theta=\partial \tau /\partial \gamma$) of high-purity single crystal Copper during torsion obtained by \citet{Gottler73_PM}. Interestingly, the model captures also the stages I and II.  The initial values and material properties are given in Table \ref{tab:parameters}.
\begin{table}
\def\arraystretch{1.5}
\begin{tabular}{ll|ll}
	 \toprule	 
	 	 
	 $G$ & $48[{\rm GPa}]$ &
	 $b$ & $0.256[{\rm nm}]$ \\ 	 
	 $\alpha$ & $0.27$ & 
	 $T$ & $0.3$ \\ 
	\rhot & $2\times 10^{-12}[{\rm m}^{-2}]$  &
	\qt & $2.86\times10^{16}[{\rm m}^{-3}]$ \\ 
	$d_{\rm ann}$ & $38 \times b$ &  \\ 	 
	$c_{\rm src}$ & $0.032$ &
	$\eta$ & $3.92$ \\
	 \bottomrule
\end{tabular} 
	\label{tab:parameters}
\caption{Material properties, initial values of dislocation densities of Copper. }
\end{table}
 The initial microstructure consists of a small density of low curvature dislocation loops which have low flow stress and line tension. This facilitates the  free flow of dislocations which is the characteristic of the first stage of work hardening (marked with (I) in \figref{fig:hardening}-top-right).  The initial growth of dislocation density is associated with expansion of dislocation loops. In this stage the curvature of microstructure w.r.t. dislocation spacing rapidly increases which indicates that dislocations become more and more entangled. This can be parametrized by the variable $\Phi=\qt/(\rhot)^{1.5}$ as depicted by  \Figref{fig:hardening}-bottom-right. As density increases, the dynamic sources become more prominent. New dislocation loops are generated and the curvature of the system increases. In the second stage, the hardening rate $\partial\tau/\partial\gamma$ reaches its maximum around $\tau=G/120$. The growth rate of dislocation density decreases which indicates the start of dynamic recovery through recombination of dislocations. In the third stage, the hardening rate decreases monotonically as dislocation density saturates. The late stages of hardening (IV, V)  exhibit themselves as a plateau at the end of the hardening rate plot and are commonly associated with dislocation cell formation. Therefore \CDDO\, cannot capture these stages. To capture these stages, one might need to use higher order non local models such as \CDDI\, and \CDDII\, which are cable of accounting for dislocation transport and capture cell formation \citep{Sandfeld15_MSMSEa}.

In our treatment we have neglected the slip contribution of segments that move on the cross slip plane during cross-slip induced recombination processes. We are now in a position to estimate this contribution, which we showed to be of the order of $f_{\rm cs} \approx q y_{\rm cs}/(2 \pi \rho)$ relative to the amount of slip on the primary slip plane. An upper estimate of the cross slip height $y_{\rm cs}$ leading to a recombination process is provided by the dislocation spacing. Hence, $f_{\rm cs} \approx \Phi/(2 \pi) \le 0.013$ at all strains considered. We conclude that in standard work hardening processes this contribution is negligible. 

\section{Summary and Conclusion}
We revisited the continuum dislocation dynamics (CDD) theory which describes conservative motion of dislocations in terms of series of hierarchical evolution equations of dislocation alignment tensors.  Unlike theories based on the Kröner-Nye tensor which measures the excess dislocation density, in CDD, dislocations of different orientation can coexist within an elementary volume. Due to this fundamental difference, in CDD, dislocations interactions should be dealt with a different approach than in GND-based theories. We introduced models for climb and cross-slip  annihilation mechanisms. The annihilation rates of alignment tensors for the first and the second order CDD theories \CDDI\, and \CDDII\, were calculated in \appref{app:climb}  and \appref{app:cross}. Later we discussed models for incorporating the activation of Frank-Read, double cross slip and glissile junction sources into CDD theory. Due to the dynamic nature of source mechanisms, ensembles of DDD simulations are needed to characterize the correlation matrices which emerge in the continuum formulation of these mechanisms. 
We outline the structure of the first and second order CDD theories with annihilation and sources in \appref{app:CDDI} and \appref{app:CDDII} respectively. We finally demonstrated that by including annihilation and generation mechanism in CDD theory, even zeroth-order CDD theory (\CDDO) obtained by truncating the evolution equations at scalar level, can describe the first 3 stages of work hardening. 

\begin{backmatter}
\section*{List of Abbreviation}	
\begin{tabular}{ll}
	CCT &  classical continuum theory of dislocation \\ 
	CDD & continuum dislocation dynamics \\ 
	DDD& discrete dislocation dynamics \\ 
	DODF& dislocation orientation distribution functions \\ 
	GND & geometrically necessary dislocations\\	
	MIEP& maximum information entropy principle  \\ 
	MRSS&maximum resolved shear stress \\
	SODT & second-order dislocation density tensor 
\end{tabular} 
\section*{Declarations}
\subsection*{Competing interests}
  The authors declare that they have no competing interests.
\subsection*{Author's contributions}
M.M. developed the proposed annihilation and generation models, implemented the simulation model, and prepared the manuscript. M.Z. analysed and corrected the models, designed the numerical example and extensively edited the manuscript. Both authors read and approved the final manuscript.
\subsection*{Funding}
The authors acknowledge funding by DFG under Grant no. 1 Za 171-7/1. M.Z. also acknowledges support by the Chinese government under the Program for the Introduction of Renowned Overseas Professors (MS2016XNJT044). 
\subsection*{Acknowledgements}
Not applicable
\subsection*{Availability of data and material}
Not applicable

\bibliographystyle{bmc-mathphys} 
\bibliography{bmc_article}      
%
\nocite{label}
\newpage
\appendix
\section{Approximating the DODF using Maximum Information Entropy Principle}\label{app:MIEP}
\citet{Monavari16_JMPS}  proposed using the  Maximum Information Entropy Principle (MIEP) to derive closure approximations for infinite hierarchy of CDD evolution equations. The fundamental idea is to estimate the DODF based upon the information contained in alignment tensors up to order $k$, and then use the estimated DODF to evaluate, from Eq. (\ref{eq:An}), the missing alignment tensor $\Brho^{(k+1)}$. By using the method of Lagrange multipliers, we can construct a DODF which has maximum information entropy and is consistent with the known alignment tensors. The CDD theory constructed by using this DODF to estimate $\Brho^{(k+1)}$ and thus obtain a closed set of equations is called the k-th order CDD theory (${\rm CDD}^{(k)}$). We can reduce the number of unknowns by assuming that the reconstructed DODF is symmetric around GND direction $\vp_{\Brho}=\tan^{-1}{(\frac{\lktwo}{\lkone})}$ and rotate the coordinates such that the GND vector becomes parallel to $x$ direction. In this case the DODF takes the form: 
\begin{align}\label{eq:DODF_sym}
p(\vp)&=\frac{1}{Z}\exp\left[-\sum_{i=1}^k \lambda_i \cos^i (\vp     -\vp_{\Brho} )\right]
\end{align}
  where the partition function of the distributions is: 
\begin{align}
\label{eq:z_sym}
Z&= \oint \,\!\! \exp\left(-\sum_{i=1}^{n} \lambda_i \cos^i(\vp     -\vp_{\Brho} ) \right)\td \vp,
\end{align} 
and $\lambda_{i}$ are the Lagrangian multipliers which are functions of known alignment tensors. 
We obtain the DODF of \CDDI\, and \CDDII\, by truncating the \eqref{eq:DODF_sym} at the first and the second order respectively: 
\begin{align}\label{eq:CDDI_DODF}
\CDDI:\quad p(\vp)&=\frac{1}{Z}\exp(-\lambda_{1}\cos(\vp-\vp_{\Brho}))\\
\label{eq:CDDII_DODF}
\CDDII:\quad p(\vp)&=\frac{1}{Z}\exp(-\lambda_{1}\cos(\vp-\vp_{\Brho})-\lambda_{2}\cos^2(\vp-\vp_{\Brho}))
\end{align}
The Lagrangian multipliers can be expressed as functions of dislocation moments $M^{(k)}$ which we define as the first components of the alignment tensors in the rotated coordinates: $M^{(k)}:=\rho'^{(k)}_{1\dots1}=\rho^{(k)}_{1\dots1}(\vp-\vp_\Brho)$. For instance, the first moment is the ratio of GND density to total density and the second moment describes the average distribution of density w.r.t GND:
\begin{align}
\MI&=|\AI|/\AO,\\
\MII&=\left(\AIIoneone\lkone\lkone+2\AIIonetwo\lkone\lktwo+\AIItwotwo\lktwo\lktwo\right)/\rhot.
\end{align}
The alignment tensor series can also be expressed in terms of moments functions:
\begin{align}
	\AI/\rhot&= \MI\\
	\AII/\rhot&=\MII \lk\otimes\lk + (1-\MII) \lkp\otimes\lkp ,\dots
\end{align}
The higher order moment functions and consequently the alignment tensors can be estimated using the reconstructed DODF. 
\section{Climb annihilation in \CDDI\,and \CDDII}\label{app:climb}
In order to find the climb annihilation rate of the the alignment tensors in \CDDI\, and \CDDII\, first we find the annihilation rate of the moment functions:
\begin{align}
\dot \Brho'^{(k)}_{1\dots 1}|_{\rm cb} &= -4 y_{\rm cb} v \rhot  \rhot  f^{(k)}_{\rm cb}(\lambda_{1},\vp_{\Brho}),
\end{align}
where $\Brho'^{(k)}_{1\dots 1}=\rhot M^{(k)}$ is the first component of the k-th order alignment tensor in the rotated coordinate system. $f^{(k)}_{\rm cb}$ is the climb annihilation function of order $k$: 
\begin{align}
f^{(k)}_{\rm cb} &=  \oint p(\vp)\left[  \int_{\vp -\frac{\pi}{2}}^{\vp+ \frac{\pi}{2}} p(\pi+2\theta-\vp) \cos^2(\vp-\theta) (\cos^k(\vp)-\frac{|\Bs''|}{2}(\Bl''_1)^{k}) {\rm d}\theta  \right]   {\rm d}\vp\\
&= \oint p(\vp)\left[  \int_{\vp -\frac{\pi}{2}}^{\vp+ \frac{\pi}{2}} p(\pi+2\theta-\vp) \cos^2(\vp-\theta) (\cos^k(\vp)-\frac{(\Bs''_1)^k}{2(|\Bs''|) ^{k-1}}) {\rm d}\theta  \right]   {\rm d}\vp\label{eq:Ann_f_all}\\
\dot \rho^{(k)}_{\rm cb} &= -4 y_{\rm cb} v \rho^2 \oint \oint \Theta(\Delta \vp - |\vp + \vp' - \pi|) 
\cos^2(\vp) \left[\Bl^{(k)}(\vp) - \left|\sin(\vp)\right| \Bl^{(k)}(\pi/2)\right]  {\rm d}\vp'{\rm d}\vp\nonumber\\
&-4 y_{\rm cb} v \rho^2 \oint \oint \Theta(\Delta \vp - |\vp + \vp' - 3\pi|) 
\cos^2(\vp) \left[\Bl^{(k)}(\vp) - \left|\sin(\vp)\right| \Bl^{(k)}(3\pi/2)\right]  {\rm d}\vp'{\rm d}\vp.
\end{align}

Using these relation we obtain the annihilation rate of $\rhot$ as:
\begin{align}
\dot \rho_{\rm cb} &= -4 y_{\rm cb} v \rhot  \rhot  f^{(0)}_{\rm cb}(\lambda_{1},\vp_{\Brho}),
\label{eq:Ann_climb_rhot}
\end{align}
where  $f^{(0)}_{\rm cb}(\lambda_{1},\vp_{\Brho})$ is the zeroth-order climb annihilation function defined as:
\begin{align}
f^{(0)}_{\rm cb}(\lambda_{1},\vp_{\Brho}) &=      \frac{1}{Z^2}\oint\exp(-\lambda_{1}\cos(\vp-\vp_{\Brho}))  \\\nonumber &\times\left[  \int_{\vp -\frac{\pi}{2}}^{\vp+ \frac{\pi}{2}}\exp(-\lambda_{1}\cos(\pi+2\theta-\vp-\vp_{\Brho}))  \cos^2(\vp-\theta)  (1-\frac{|\Bs''|}{2}){\rm d}\theta   \right]   {\rm d}\vp.
\label{eq:Ann_climb_f0}
\end{align}
Given that the DODF of \CDDI\ is symmetric around the GND angle  $\vp_\Brho$, $f^{(0)}_{\rm cb}$ can be derived as a function of the only Lagrangian multiplier $\lambda_{1}$. It is  more physically intuitive to express this rate as a function of the corresponding first dislocation moment $\MI = |\AI|/\rhot$, which can be understood as the GND fraction of the total dislocation density. 
As depicted in \figref{fig:ann_climb_CDDI},
$f^{(0)}_{\rm cb}$ does not correspond to the parabolic rate expected by bimolecular annihilation of straight dislocation lines. 
The annihilation of  \rhot\, can be approximated by:
\begin{align}
\dot \rhot_{\rm cb} &= - 1.2 y_{\rm cb} v \rhot \rhot (1-1.5(\MI)^2+0.5(\MI)^6).
\end{align}  
\begin{figure}[ht]
	\centering
	\includegraphics[width=0.6\linewidth]{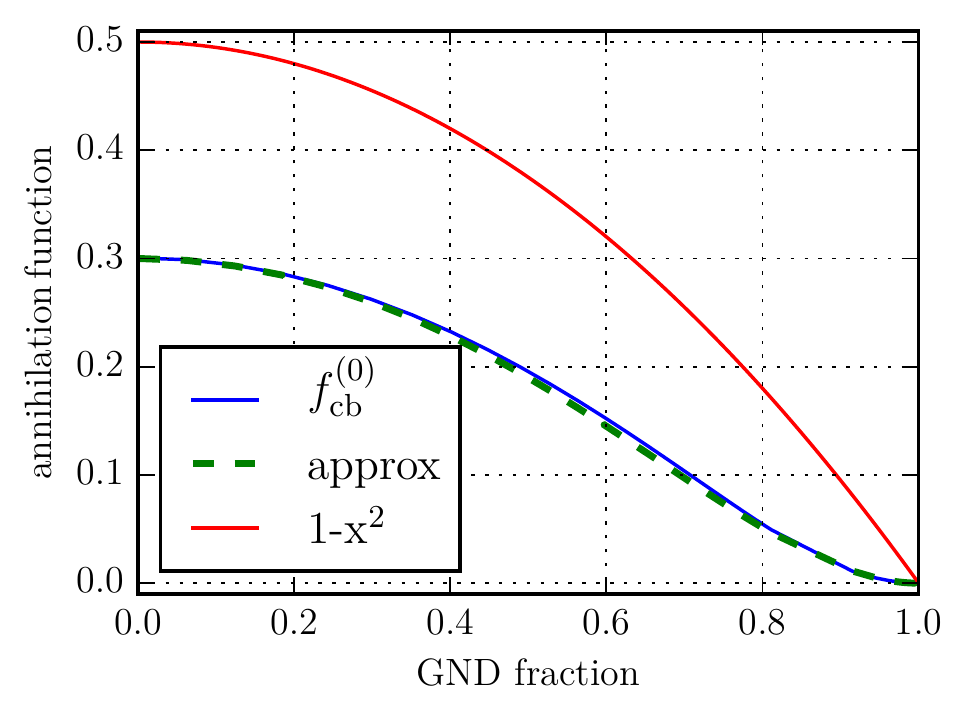}
	\caption[Climb annihilation rate of \rhot]{Blue line: climb annihilation function of \rhot\, as a function of the GND fraction. Green dashed-line: analytical fit ($0.3(1-1.5x^2+0.5x^6)$) to the  annihilation rate. Red line: parabolic rate ($0.5(1-x^2)$) predicted by bimolecular annihilation.}
	\label{fig:ann_climb_CDDI}
\end{figure}  
\begin{figure}[t]
	\centering
	\includegraphics[width=0.99\linewidth]{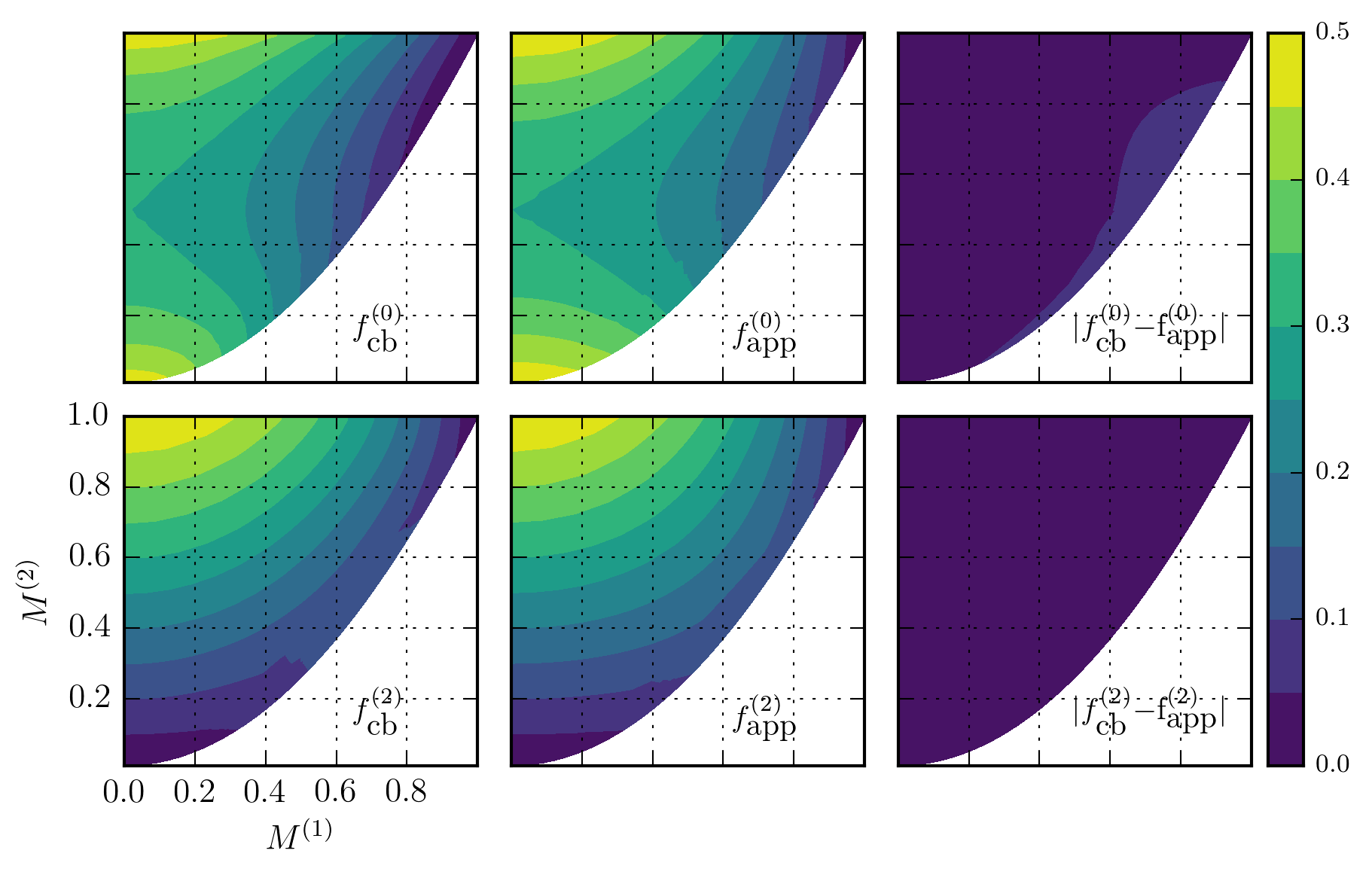}
	\caption[Climb annihilation functions]{Left column: the Zeroth and the second order annihilation functions as functions of \MI\, and  \MII. Center column: polynomial approximations of the annihilation functions. Right column: absolute errors of the approximations. }
	\label{fig:anil_climb_CDDII}
\end{figure}

Similar to the \CDDI, the annihilation rate of the first three moment function of \CDDII\,can be calculated using its DODF:
\begin{align}
\left.\dot \rhot\right|_{\rm cb} &= 4 y_{\rm cb} v \rhot\rhot f^{(0)}_{\rm cb}(\MI,\MII),\\
\left.\dot \Brho'^{(1)}_{1}\right|_{\rm cb} &= 4 y_{\rm cb} v \rhot f^{(1)}_{\rm cb}(\MI,\MII)=0,\\
\left.\dot \Brho'^{(2)}_{11}\right|_{\rm cb} &= 4 y_{\rm cb} v \rhot f^{(2)}_{\rm cb}(\MI,\MII).
\label{eq:Ann_rhot_all}
\end{align}

Note that the first order annihilation function  is always zero by definition ($f^{(1)}_{\rm cb}=0$). 
\Figref{fig:anil_climb_CDDII} depicts the zeroth and second order annihilation functions and their analytical approximation as  functions of \MI\, and \MII:
\begin{align}
f^{(0)}_{\rm cb}(\MI,\MII) &\approx  (0.8(\MII-.5)^2+0.3) (1-(\MI)^2),\\
f^{(2)}_{\rm cb}(\MI,\MII) &\approx 0.5 \MII (1-(\MI)^2).
\end{align}
In the limit case of $\MII=1$, where dislocations become parallel straight lines, annihilation functions converge to parabolic bi-molecular annihilation. 
The annihilation rate  of \AII\, can be evaluated using the relation between moment functions  and alignment tensors given by \citet{Monavari16_JMPS}:
\begin{align}\label{eq:Ann_climb_AII}
\dot \Brho^{(2)}_{\rm cb} &=   \dot \Brho'^{(2)}_{11}|_{\rm cb} \lk\otimes\lk + (\dot \rho_{\rm cb} -\dot \Brho'^{(2)}_{11}|_{\rm cb} ) \lkp\otimes\lkp\\ \nonumber &=-4 y_{\rm cb} v \rhot \rhot \left[ f^{(2)}_{\rm cb} \lk\otimes\lk + (f^{(0)}_{\rm cb}-f^{(2)}_{\rm cb}) \lkp\otimes\lkp\right].
\end{align}
Assuming an equi-convex microstructure where all dislocations have the same (mean) curvature, the annihilation rate of the total curvature density can be straightforwardly evaluated from the dislocation density annihilation rate:
\begin{align}
\label{eq:Ann_climb_qt}
\dot \qt _{\rm cb}  &=  \dot \rhot_{\rm cb} \frac{\qt}{\rhot}. 
\end{align}    
The concomitant reduction in dislocation curvature density decreases the elongation (source) term $v\qt$ in the evolution equation of the total dislocation density \eqref{eq:dA0dt} -- an effect which has an important long-term impact on the evolution of the dislocation  microstructure and may outweigh the direct effect of annihilation. 
The total annihilation rate is the sum of annihilation by cross slip and climb mechanisms. 
\section{Cross slip annihilation in \CDDI\, and \CDDII}  \label{app:cross}

\subsection[Cross slip annihilation in \CDDI]{Cross slip annihilation in \CDDI}  
The cross slip annihilation rate of DODF in \CDDI\,can be calculated by plugging the DODF of \CDDI\,given by \eqref{eq:CDDI_DODF} into \eqref{eq:Ann_rhok_cross}. Assuming that the dislocations have a smooth angular distribution which can be approximated as constant over the small angle interval $2\Delta \vp$,  \eqref{eq:Ann_rhok_cross}  can be further simplified:
\begin{align}\label{eq:ann_cross_DODF_CDDI}
\dot \rho_{\rm cs}(\vp) &=  -8 \Delta \vp  y_{\rm cs} v \rho(\vp) \rho(\pi - \vp) \cos^2 (\vp) (1-|\sin (\vp)|)   \\
&= -8 \Delta \vp  y_{\rm cs} v \frac{\rhot \rhot}{Z^2} \exp(-\lambda_1 \cos (\vp-\vp_\Brho) -\lambda_1 \cos (\pi-(\vp-\vp_\Brho))) \cos^2 (\vp) (1-|\sin (\vp)|)   \nonumber
\end{align}

The annihilation rate of the zeroth order  alignment tensor (total dislocation density) is  given by integrating \eqref{eq:ann_cross_DODF_CDDI} over all orientations:
\begin{align}\label{eq:Ann_DODF_CDDI}
\dot\rho_{\rm cs}  &= -8 \Delta \vp  y_{\rm cs} v \rhot \rhot \frac{1}{Z^2} \oint  \exp(-\lambda_1 \cos (\vp-\vp_\Brho) -\lambda_1 \cos (\pi-(\vp-\vp_\Brho))) \cos^2 (\vp) (1-|\sin (\vp)|)  \td  \vp\nonumber\\ 
&=-8\Delta\vp  v y_{\rm cs} \rhot \rhot f^0_{\rm cs}.
\end{align}
$f^0_{\rm cs}$ is a function of the symmetry angle of DODF  $\vp_\Brho$ and the Lagrangian multiplier $\lambda_1$ or the corresponding $\MI$. 
We are especially interested in limit cases where the DODF is symmetric around the screw orientation and edge orientation which correspond to the axes of \Figref{fig:ann_cross_slip_CDDI}(right). 
In the first case the GND vector is aligned with the screw orientations $\vp_\Brho =0$ and $\vp_\Brho = \pi$ such that $\rho(\vp)=\rho(-\vp)$ and $\MI=\rho_1/\rhot$. Hence \eqref{eq:Ann_DODF_CDDI} becomes:
\begin{align}
\label{eq:AnnScrewGND}
\dot\rho_{\rm cs} &=-4y_{\rm cs} v(\rhot)^2 \frac{2\Delta\vp}{Z^2}  \left[\oint \cos^2 (\vp) (1-|\sin (\vp)|)  {\rm d}\vp \right] y_{\rm cs} v\nonumber \\
&=-4y_{\rm cs} v(\rhot)^2 \frac{2\Delta\vp}{Z^2}(\pi-\frac{4}{3}),
\end{align}
where $Z^2$ is a function of the first moment $\MI$.

The second case corresponds to a microstructure where $\rho(\vp)=\rho(\pi-\vp)$. Using this symmetry property and the DODF given by  \eqref{eq:CDDI_DODF}, the rate of reduction in total dislocation density in \CDDI\,can be evaluated as   
\begin{align}
\label{eq:AnnEdgeGND}
\dot\rho_{\rm cs}&= -4y_{\rm cs} v(\rhot)^2 \frac{2\Delta\vp}{Z^2}\left[\oint \exp(-2\lambda_1 \sin(\vp))\cos^2 (\vp) (1-|\sin (\vp)|) {\rm d}\vp
\right] .
\end{align}
For a completely isotropic dislocation arrangement, $\lambda_1 = 0$ and $Z = 2\pi$, we obtain in both cases:  
\begin{align}
\label{eq:AnnIsoRho}
\dot\rho_{\rm cs} &=-4 y_{\rm cs}v (\rhot)^2   (\frac{2\Delta\vp}{4\pi^2})(\pi-\frac{4}{3}).
\end{align}
\begin{figure}[t]
	\centering
	\includegraphics[width=0.6\linewidth]{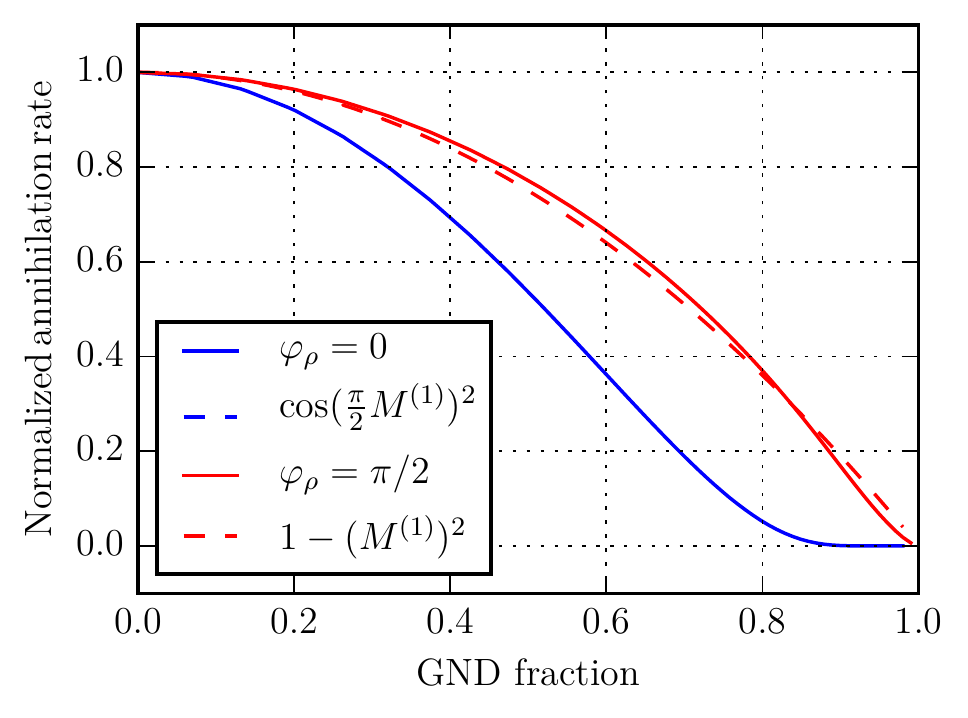}
	\caption[Normalized cross slip annihilation rate as a function of the GND fraction. ]{Normalized annihilation rate as a function of the GND fraction \MI\ for a dislocation annihilation triggered by cross slip. Blue line: normalized annihilation rate for a DODF symmetric around screw orientation ($\vp_\Brho=0,\pi$). Red line: normalized annihilation rate for a DODF symmetric around edge orientation ($\vp_\Brho=\frac{\pi}{2},\frac{3\pi}{2}$). Dashed lines: Parabolic annihilation rate expected from the kinetic theory.}
	\label{fig:annMI}
\end{figure}
\Figref{fig:annMI} compares these two limit cases with the  parabolic dependency expected according to kinetic theory for a system of straight 
parallel dislocations (dashed red line). 
In general, the annihilation rate can be approximated by interpolating between these two cases. \figref{fig:ann_cross_slip_CDDI} shows the annihilation rate, normalized by the value at $\MI = 0$, as a function of the GND fraction $\MI$ and the GND angle $\vp_\Brho$ or the corresponding screw and edge components of the normalized GND vector $\AI/\rhot$. We can see that the annihilation rate decreases monotonically with increasing GND fraction and goes to zero if all dislocations are GND. 
\begin{figure}
	\centering
	\includegraphics[width=0.99\linewidth]{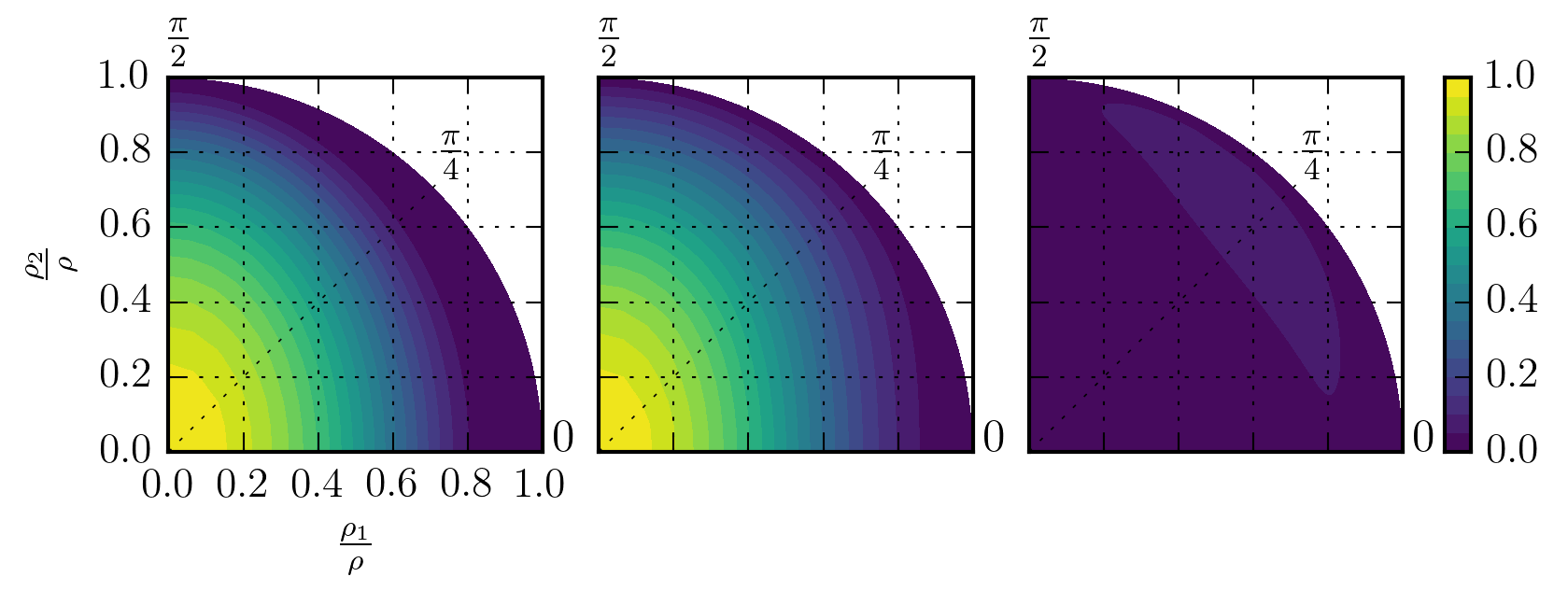}
	\caption[Cross slip annihilation function $f^0_{\rm cs}$]{Left: cross slip annihilation function $f_{\rm ann}$ of total dislocation density in $\CDDI$ plotted in polar coordinates with the first dislocation moment \MI\,as distance to the origin and the GND angle $\vp_\Brho$. The equivalent Cartesian coordinates  are the screw and edge components of the normalized GND vector $\wh{\Brho}^{(1)}=\AI/\rhot$. Middle: analytical approximation of the annihilation rate $f_{\rm cs}=(\whAIone)^2\cos^2(\frac{\pi|\AI| }{2\rhot})+(\whAItwo)^2(1-(\frac{|\AI|}{\rhot})^2)$. Right: the absolute error of the approximation.      }
	\label{fig:ann_cross_slip_CDDI}
\end{figure}

Assuming an equi-convex microstructure, the annihilation rate  of curvature density becomes:
\begin{align}
\label{eq:Ann_cross_qt}
\dot \qt _{\rm cs}  &=  \dot \rhot _{\rm cs} \frac{\qt}{\rhot}. 
\end{align}    

\subsection[Cross slip annihilation in \CDDII]{Cross slip annihilation in \CDDII}
The cross slip annihilation rate of the second order alignment tensors in \CDDII\, can be calculated by plugging the corresponding DODF into  \eqref{eq:Ann_rhok_cross}.  Assuming the symmetric DODF given by \eqref{eq:CDDII_DODF} the annihilation rate of \AII\, takes the form of::
\begin{align}\label{eq:Ann_DODF_CDDII}
\dot\Brho^{(2)}_{\rm cs}&= -4 v  y_{\rm cs} \rhot \rhot   \Bf^{(2)}_{\rm cs} (\lambda_1,\lambda_2,\vp_\Brho) 
\end{align}
where $\Bf^{(2)}_{\rm cs} $ is a  symmetric second order tensorial function of the symmetry angle $\vp_\Brho$ and the Lagrangian multipliers $\lambda_1$ and $\lambda_2$  (or their corresponding first two moment functions).  Each component of  $\Bf^{(2)}_{\rm cs} $ can be approximated by 3 dimensional tables (or 4 dimensional in case of full DODF). \Figref{fig:anil_CDDII_0} and \Figref{fig:anil_CDDII_1} depict two slice of the 3D annihilation tables  of \rhot\, and  \AII\ which correspond to symmetric DODFs around screw ($\vp_\Brho=0$) and edge ($\vp_\Brho=\frac{\pi}{2}$) orientations respectively.   For the corresponding orientation interval we use the value $\Delta \vp = \pm 15^\circ$ given by \cite{Hussein15_AM}. 
\begin{figure}
	\centering
	\includegraphics[width=0.99\linewidth]{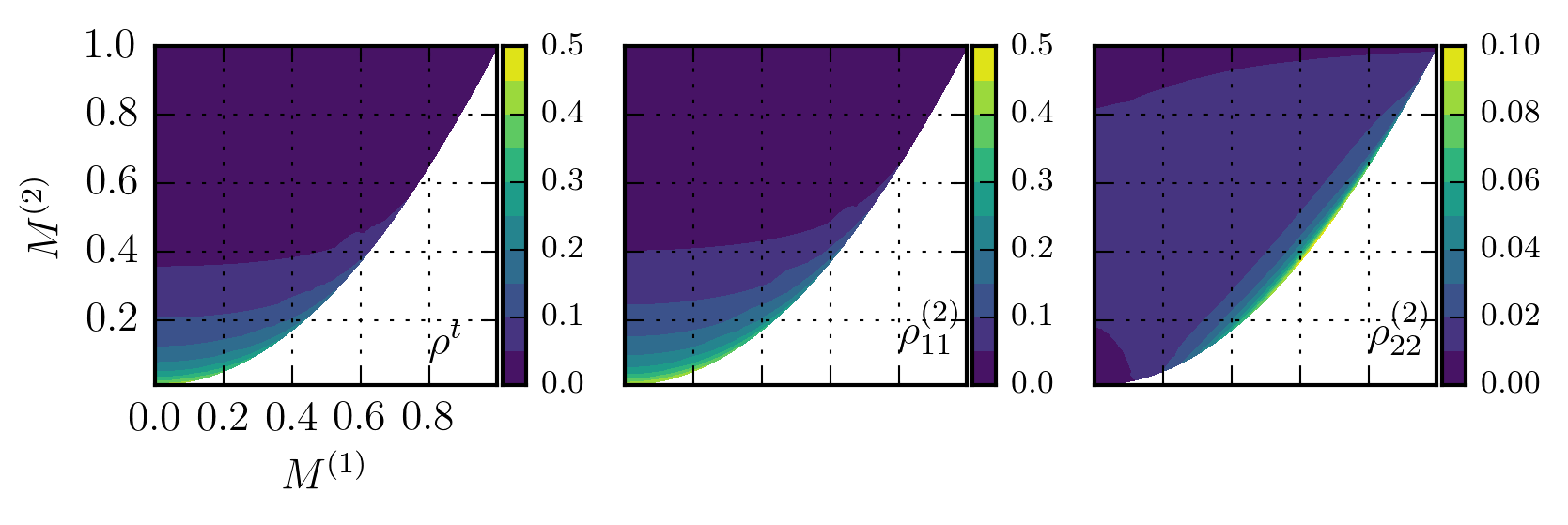}
	\caption[Cross slip annihilation rate of \CDDII\, ($\vp_\Brho=0$)]{Cross slip annihilation functions of \rhot, \AIIoneone and \AIItwotwo in  \CDDII\, for a DODF symmetric around screw orientation ($\vp_\Brho=0$). For this symmetry angle $\MI=\rho_{1}/\rhot$ and $\MII=\rho_{11}/\rhot$. Bimolecular annihilation corresponds to the upper limit of $\MII=1$. }
	\label{fig:anil_CDDII_0}
\end{figure}
\begin{figure}
	\centering
	\includegraphics[width=0.99\linewidth]{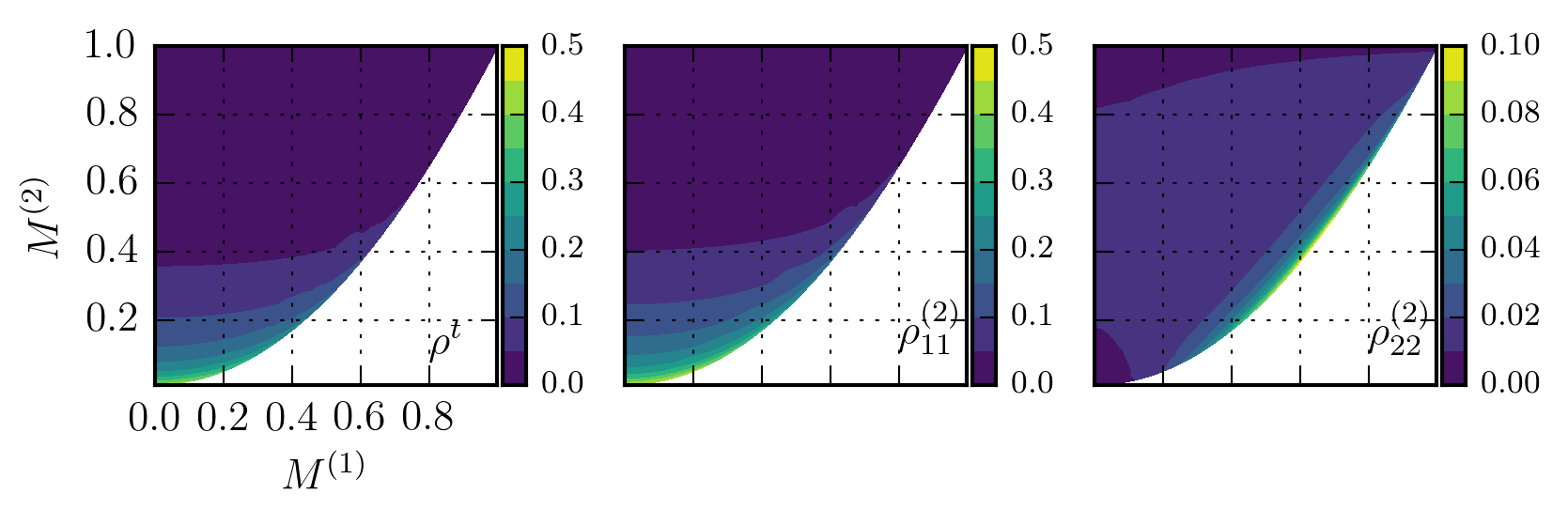}
	\caption[Cross slip annihilation rate of \CDDII\, ($\vp_\Brho=\frac{\pi}{2}$)]{Cross slip annihilation functions of \rhot, \AIIoneone and \AIItwotwo in  \CDDII\, for a DODF symmetric around edge orientation ($\vp_\Brho=\frac{\pi}{2}$). For this symmetry angle $\MI=\rho_2/\rhot$ and $\MII=\rho_{22}/\rhot$.}
	\label{fig:anil_CDDII_1}
\end{figure}

\Figref{fig:anil_CDDII_0} shows that, in the limiting cases where all dislocations are screw oriented, i.e.  $\MII=1$ and $\rho(\vp) = \rho_+ \delta(\vp) + \rho_- \delta(\pi - \vp)$, the annihilation rate follows as 
\begin{align}
\left.\dot \rho_+\right|_{\rm cs} = \left.\dot \rho_-\right|_{\rm cs} = 
- 4 \rho_+ \rho_- y_{\rm cs} v ,
\end{align}
which is the result expected by kinetic theory for particles moving in a 2D space with velocity $v$ in opposite directions and annihilating if they pass within a reaction cross-section $2 y_{\rm cs}$. 

\section{Evolution equations of \CDDI}\label{app:CDDI}
The total dislocation density $\rhot$, the dislocation density vector  $\AI$, and the total curvature density $\qt$ are the kinematic variables of \CDDI. 
In order to reconstruct the DODF and approximate $\AII$, first we have to calculate the average line direction $\lk$, the symmetry angle $\vp_\Brho$ and the first moment function \MI:
\begin{align}
\lk &= \AI/|\AI| =[\lkone,\lktwo]=[\cos(\vp_{\Brho}),\sin(\vp_{\Brho})],\\
\vp_{\Brho}&=\tan^{-1}{(\frac{\lktwo}{\lkone})},\\
\MI&=|\AI|/\AO.
\end{align}
We also remind that operator $\wh{(\bullet)}$ normalizes quantities with $\rho$; e.g. $\wh\rho_1=\frac{\AIone}{\rho}$. 

\MII\, and \AII can be approximated as \citep{Monavari16_JMPS}:
\begin{align}\label{eq:app4:M2}
\MII &\approx [2 + (\MI)^2 + (\MI)^6]/4.\\
\AII &\approx \rhot\left[ \MII \lk \otimes \lk + (1-\MII )\lkp \otimes \lkp\right]\\\nonumber
&= \rhot\left[ \MII 
\begin{bmatrix}
\lkone^2 & \lkone\lktwo  \\ \lkone\lktwo & \lktwo^2  
\end{bmatrix} 
+ (1-\MII )
\begin{bmatrix}
\lktwo^2 & -\lkone\lktwo  \\ -\lkone\lktwo & \lkone^2  
\end{bmatrix} 
\right].
\end{align}
The curvature density vector is approximated using the equi-convex assumption:
\begin{align} 
\BQ^{(1)} =- (\AI)^{\perp} \frac{\qt}{\rhot}.
\end{align}
The cross slip annihilation rate of $\rhot$ is a function of $\MI$, $\lk$ and cross slip distance $y_{\rm cs}$ \eqref{eq:Ann_DODF_CDDI}:
\begin{align}
\dot\rho_{\rm cs}  &= - v y_{\rm cs} \rhot \rhot f^0_{\rm cs}(\frac{1}{6} -\frac{4}{3\pi}) ,
\end{align}
where $f_{\rm cs}=(\whAIone)^2\cos^2(\frac{\pi|\AI| }{2\rhot})+(\whAItwo)^2(1-(\frac{|\AI|}{\rhot})^2)$.
The climb annihilation rate of $\rhot$ is a function of $\MI$ and climb distance $y_{\rm cs}$:
\begin{align}
\dot\rho_{\rm cb}  &= - 4 y_{\rm cb} v \rhot \rhot f_{\rm cb},
\end{align}
with $f_{\rm cb}\approx0.3(1-1.5(\MI)^2+0.5(\MI)^6)$.
The curvature generation rates attributed to the activation of Frank-Read sources, cross slip sources, and glissile junctions are
\begin{align}
\dot q_{\rm fr}&=\frac{2\pi}{5} v \rho_{\rm FR}^2,\\
\label{eq:qt_dcs_source_rate}
\dot q_{\rm dcs}&= \pi\frac{f_{\rm dcs}}{\eta} v \rho_{\rm s} \rhotot,\\
\dot q_{\rm gj}&=\sum_{\varsigma'}\sum_{\varsigma''}\pi f^{{\varsigma'}{\varsigma''}}_{\rm gj} v  \frac{\rho^{\varsigma'}\rho^{\varsigma''}}{\eta},\\
\end{align}
where $\rho_{\rm fr}$ is the density of the dislocation segments acting as static Frank-Read sources, $f_{\rm dcs}$ and $f_{\rm gj}$ are a correlation matrices that relate  the dislocation densities to activation of cross slip sources and glissile junctions on the considered slip system. Like the cross slip annihilation rate, the screw dislocation density $\rho_{\rm s}$ is a function of $\rhot$ and $\AI$ and can be estimated as 
\begin{align} 
\rho_{\rm s}&\approx \frac{1}{6}+\frac{5}{6((\rho_1)^2+(\rho_2)^2)}(1.1(\whAIone)^6-1.4(\whAIone)^8+1.3(\whAIone)^{14}-.16(\whAItwo)^4-.22(\whAItwo)^6+.18(\whAItwo)^8). \label{eq:cdd1screw}
\end{align}
\begin{figure}
	\centering
	\includegraphics[width=0.99\linewidth]{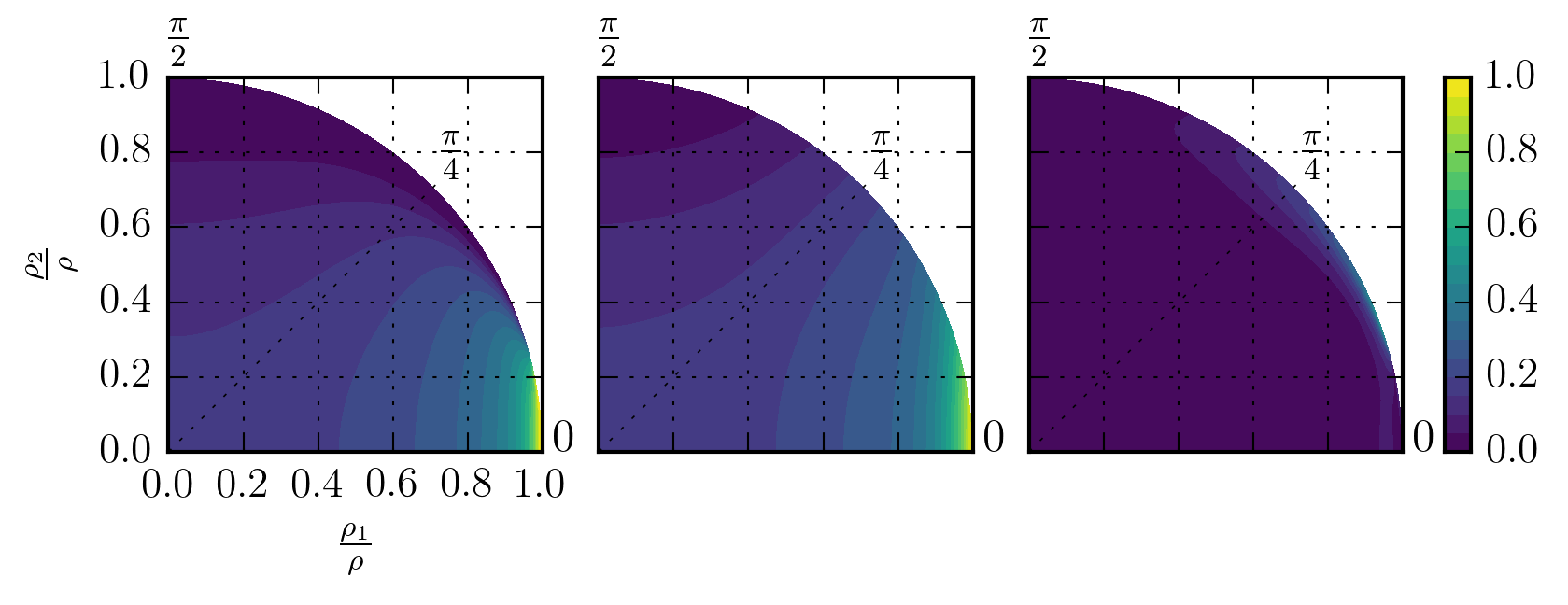}
	\caption{Ratio of screw dislocation density $\rho_s/\rho$ as a function of GND vector and total dislocation density. Left: Evaluated from integrating the DODF of \CDDI using \eqref{eq:intscrew}; Middle: Analytical approximation of screw density ratio  given by \eqref{eq:cdd1screw}; Right: the absolute error of the estimation.}
	\label{fig:cdd1screw}
\end{figure}
The total annihilation and source rates of \rhot\, and \qt\, then become:
\begin{align}
\dot\rho_{\rm ann}  &= \dot\rho_{\rm cs} +\dot\rho_{\rm cb},\\
\dot q_{\rm ann}  &= \frac{\qt}{\rhot}\dot\rho_{\rm ann},\\
\dot q_{\rm src}  &= \dot q_{\rm fr} +\dot q_{\rm dcs} +\dot q_{\rm gj}\label{eq:app:qt_sources}, \\
\end{align}
We note that source activation and annihilation do not change the GND vector $\AI$. The evolution equations for $\rhot$, $\AI$, and $\qt$ then take the form:
\begin{align}
\label{eq:app:drhotdt_CDD1}
\dot\rhot &=\nabla \cdot(v \Bve\cdot\AI)+v\qt + \dot\rho_{\rm ann} ,\\
\label{eq:app:dA1dt_CDD1}
\dot\Brho^{(1)} &= -\Bve\cdot\nabla(\rho v),\\
\label{eq:app:dqdt_CDD1}
\dot \qt &=\nabla \cdot(v\BQ^{(1)} -  \Brho^{(2)}\cdot\nabla v ) +\dot q_{\rm src}+\dot q_{\rm ann},\\
\dot{\gamma}&=\rho v b 
\end{align}
The only missing parameters of this system of equations are the correlation matrices.
\section{Evolution equations of \CDDII}\label{app:CDDII}
\CDDII\, is constructed by following the evolution of  \AII\, in addition to $\AI$ and \qt. Similar to \CDDI, first we  calculate the average line direction $\lk$, the symmetry angle $\vp_\Brho$ and the first two moment function \MI\, and  \MII:
\begin{align}
\lk &= \AI/|\AI| =[\lkone,\lktwo]=[\cos(\vp_{\Brho}),\sin(\vp_{\Brho})],\\
\vp_{\Brho}&=\tan^{-1}{(\frac{\lktwo}{\lkone})},\\
\MI&=|\AI|/\AO,\\
\MII&=\left(\AIIoneone\lkone\lkone+2\AIIonetwo\lkone\lktwo+\AIItwotwo\lktwo\lktwo\right)/\rhot.
\end{align}
$\AIII$ is then given by  approximated as:
\begin{align}
\label{eq:app:A3_symbolic}
\AIII/\rhot=&\MIII  \lk\otimes\lk\otimes\lk\\
\nonumber &+\left(\MI-\MIII\right )(\lk\otimes\lkp\otimes\lkp
+\lkp\otimes\lk\otimes\lkp
+\lkp\otimes\lkp\otimes\lk), \end{align}
where the third order moment function \MIII\, is approximated as $\MIII\approx \MI\sqrt{\MII}$. 
The curvature density vector is given by the divergence of \AII:
\begin{align}
\label{eq:QII_equi}
\QI &=\nabla \cdot \AII.
\end{align}
The second order auxiliary curvature density becomes: 
\begin{align}
\QII&= \frac{\qt}{2|\QI|^2}\left[ (1+\Phi) {\QI} \otimes {\QI} + (1-\Phi){\QI}^{\perp} \otimes {\QI}^{\perp}\right] ,
\end{align}
where $\Phi \approx (|\QI|/\qt)^2(1+(|\QI|/\qt)^4)/2$.

\eqref{eq:Ann_DODF_CDDII} gives the  cross slip annihilation rate of $\AII$: 
\begin{align}
\dot\Brho^{(2)}_{\rm cs}&= -4 v  y_{\rm cs} \rhot \rhot   \Bf^{(2)}_{\rm cs} (\lambda_1,\lambda_2,\vp_\Brho) ,
\end{align}
where $\Bf_{\rm cs}$ is a tensorial function of $\MI$, $\MII$ and $\lk$ and can be tabulated numerically.
The climb annihilation rate of $\AII$ is given by \eqref{eq:Ann_climb_AII}:
\begin{align}
\dot\Brho^{(2)}_{\rm cb} &= -4 y_{\rm cb} v \rhot \rhot \left[ f^{(2)}_{\rm cb} \lk\otimes\lk + (f^{(0)}_{\rm cb}-f^{(2)}_{\rm cb}) \lkp\otimes\lkp\right],
\end{align}

where the zeroth and the  second order climb annihilation functions are approximated as:
\begin{align}
f^{(0)}_{\rm cb}(\MI,\MII) &\approx  (0.8(\MII-.5)^2+0.3) (1-(\MI)^2),\\
f^{(2)}_{\rm cb}(\MI,\MII) &\approx 0.5 \MII (1-(\MI)^2).
\end{align}

Total annihilation rate of $\AII$ is given by the summation of the cross slip and the climb annihilation rates $\dot\Brho^{(2)}_{\rm ann}  = \dot\Brho^{(2)}_{\rm cs} +\dot\Brho^{(2)}_{\rm cb}$.

In \CDDII,  density of screw dislocations can be approximated as $\rho_s\approx\rho_{11}$.
Similar to \CDDI, the contribution of dynamic sources to $q$ can be calculated from \eqref{eq:app:qt_sources}. 

The evolution equations for $\AI$,  $\AII$ and $\qt$ then take the form:
\begin{align}
\label{eq:app:dkappadt_A2}
\dot\Brho^{(1)}&= -\Bve\cdot\nabla(\rho v)\\
\label{eq:app:dA2dt}
\dot\Brho^{(2)}&=\left[-\Bve\cdot\nabla   (v  \AI)+v\QII-\Bve \cdot \AIII\cdot\nabla v \right]_{{\rm sym}}+\dot\Brho^{(2)}_{\rm ann},\\
\label{eq:app:dqtdt_A2}
\dot\qt&=\nabla \cdot ( v\BQ^{(1)} - \AII\cdot \nabla v )+\dot q_{\rm src}+\dot q_{\rm ann},\\
\dot{\gamma}&={\rm Tr}(\AII) v b 
\end{align}

\end{backmatter}
\end{document}